\documentclass[prb,twocolumn,notitlepage,preprintnumbers,amsmath,amssymb,superscriptaddress]{revtex4-2}
\usepackage{graphicx,bm,amsmath}
\usepackage{amsmath,amssymb}
\usepackage{graphicx}
\usepackage{wasysym}
\usepackage{amsfonts}
\usepackage{bm}
\usepackage{enumerate}
\usepackage{color}
\usepackage{xcolor}
\usepackage{epstopdf}
\usepackage{latexsym}
\usepackage[breaklinks,colorlinks = true,linkcolor = red,urlcolor=cyan,citecolor=red]{hyperref}
\usepackage{stmaryrd} 
\usepackage{braket} 
\usepackage[textsize=tiny]{todonotes} 

\newcommand{\br}{{\bf r}}
\newcommand{\bR}{{\bf R}}
\newcommand{\bk}{{\bf k}}
\newcommand{\bK}{{\bf K}}

\usepackage[normalem]{ulem}

\begin{document}
\title{Connecting the avoided quantum critical point to the magic-angle transition in three-dimensional Weyl semimetals}

\author{J. H. Pixley}
\affiliation{Department of Physics and Astronomy, Center for Materials Theory, Rutgers University, Piscataway, NJ 08854 USA}
\affiliation{Center for Computational Quantum Physics, Flatiron Institute, 162 5th Avenue, New York, NY 10010} 
\author{David A. Huse}
\affiliation{Department of Physics, Princeton University, Princeton, New Jersey 08544, USA}
\author{Justin H. Wilson}
\affiliation{Department of Physics and Astronomy, and Center for Computation and Technology, Louisiana State University, Baton Rouge, LA 70803, USA}
\date{\today}

\begin{abstract}
We theoretically study the interplay of short-ranged random and quasiperiodic static potentials on the low-energy properties of three-dimensional Weyl semimetals. 
This setting allows us to investigate the connection between the semimetal to diffusive metal ``magic-angle'' phase transition due to quasiperiodicity and the rare-region induced crossover at an avoided quantum critical point (AQCP) due to disorder. 
We show that in the presence of both random and quasiperiodic potentials the AQCP becomes lines of crossovers, which terminate at magic-angle critical points in the quasiperiodic, disorder-free limit. 
We analyze the magic-angle transition by approaching it along these lines of avoided transitions, which unveils a rich miniband structure and several AQCPs. 
These effects can be witnessed in cold-atomic experiments through potential engineering on semimetallic band structures. 
\end{abstract}

\maketitle

\section{Introduction}
There is a significant push to discover and understand the nature of gapless topological materials. This has been fueled by 
the experimental discovery of three-dimensional (3D) topological Dirac and Weyl semimetals in weakly correlated narrow gap semiconductors \cite{borisenkoExperimentalRealizationThreedimensional2014,liuDiscoveryThreedimensionalTopological2014,Neupane-2014,lvExperimentalDiscoveryWeyl2015,lv_observation_2015,Xu3-2015,Xu2-2015,xuExperimentalDiscoveryTopological2015,xuObservationWeylNodes2016}, as well as their observation in several strongly correlated materials \cite{tafti_pressure-tuned_2012,sushkov_optical_2015,laiWeylKondoSemimetal2018,changParityviolatingHybridizationHeavy2018,telang_anomalous_2019,dzsaberGiantSpontaneousHall2019,kurodaEvidenceMagneticWeyl2017,guoEvidenceWeylFermions2018,liu_magnetic_2021}.  
However, the Fermi energy does not typically coincide with the Weyl or Dirac touching points in the band structure, making the effects on the low-energy thermodynamic properties indirect. 
Nonetheless, nodal touching points have been identified using a combination of ARPES experiments \cite{liuDiscoveryThreedimensionalTopological2014,borisenkoExperimentalRealizationThreedimensional2014,liuStableThreedimensionalTopological2014,Neupane-2014,Xu-2015} and \textit{ab initio} calculations \cite{Huang-2015,wengWeylSemimetalPhase2015}, while their manifestation in transport arises through a negative magnetoresistance \cite{kimDiracWeylFermions2013,liangUltrahighMobilityGiant2015,liChiralMagneticEffect2016}. 
These measurements provide a systematic means to identify the existence of Dirac and Weyl nodes in several weakly correlated material candidates.
Recently, the demonstration of a 3D Weyl semimetal in an ultracold atom experiment using artificial spin-orbit coupling~\cite{wang_realization_2021} opens the door to a new level of control over Weyl semimetals. 
These systems are tunable; filling is controlled by the number of atoms in the trap, and disorder and lattice imperfections are removed altogether. 
Therefore, perturbations can be turned on at will to determine the fate of Weyl semimetals experimentally while opening the door to study effects that are out of reach in solid-state compounds.

Due to the interplay of topology and a vanishing pseudogap density of states,  single particle perturbations can have several non-trivial effects. In particular, the effects of disorder on noninteracting Weyl semimetals have been well studied~\cite{syzranovHighdimensionalDisorderdrivenPhenomena2018,pixley_rare_2021}. 
Within perturbative (e.g., self-consistent Born~\cite{shindou_effects_2009,Ominato-2014}, large-$N$~\cite{ryu_disorder-induced_2012}, and renormalization group~\cite{Goswami-2011,Sergey-2015}) treatments of the problem, a disorder-driven quantum critical point was found. 
However, when taking into account the non-perturbative effects of disorder, rare regions of the random potential give rise to power-law quasibound states where the disorder is atypically large and cannot be treated perturbatively~\cite{nandkishoreRareRegionEffects2014}. 
These rare states were found to endow the Weyl semimetal with a finite density of states at the Weyl node, destabilizing the Weyl semimetal phase into a diffusive metal for any weak random potential~\cite{pixley_rare_2021}. 
As a result, it was shown that the putative critical point is rounded out into a crossover, dubbed an avoided quantum critical point (AQCP)~\cite{Pixley-2016,PixleyBR-2016,pixleySingleparticleExcitationsDisordered2017,Guararie-2017}. 
Instanton fluctuation calculations (about the saddle point) for a single Weyl cone found that this picture is modified~\cite{Buchold-2018}, and these results were interpreted in terms of a non-trivial scattering phase shift~\cite{Buchold-B2018}. However, it was then later shown that such phase shifts are inherently problematic as their conclusions violate Levinson's Theorem~\cite{galindo2012quantum} and instead  the AQCP is the correct description~\cite{Pires-2021}. This was also consistent with a numerical study of a disordered single Weyl cone in the continuum limit that showed the transition remains strongly avoided with essentially the same kind of AQCP as previously studied lattice models with multiple Weyl cones~\cite{wilsonAvoidedQuantumCriticality2020}.

On the other hand, the fate of Weyl semimetals in the presence of quasiperiodicity that does not have any rare regions (due to the potential being infinitely long-range correlated) was only considered recently. 
It was numerically shown~\cite{Pixley-2018}, and then rigorously proven \cite{mastropietroStabilityWeylSemimetals2020} that the Weyl semimetal phase is stable to a quasiperiodic potential. 
As a result, quasiperiodicity drives a bona fide semimetal to diffusive metal phase transition at a non-zero critical quasiperiodic strength~\cite{Pixley-2018}. 
At this transition the Weyl velocity goes to zero continuously, the density of states becomes non-analytic, and the single particle wavefunctions at the Dirac node energy delocalize in momentum space. 
Studies of similar effects in two-dimensional Dirac semimetal models~\cite{fuMagicangleSemimetals2020,chou_magic-angle_2020,Fu-2021} have linked this quantum phase transition with the magic-angle phenomena originally discovered in twisted bilayer graphene~\cite{bistritzer2011moire} (and extended to incorporate incommensurate effects~\cite{fuMagicangleSemimetals2020,gonccalves2021incommensurability}). 
Thus, the transition that was originally sought in disordered Weyl semimetals was uncovered in the quasiperiodic limit by removing rare regions from the problem.
We therefore refer to the 
critical point due to a quasiperiodic potential as a ``magic-angle transition'' (MAT); here the ``angle'' refers to the incommensurate wave vector characterizing the quasiperiodic potential.

In the following manuscript, we make a direct link between the avoided transition and the magic-angle quantum critical point in 3D. 
We do so by considering how the avoided transition is connected to the magic-angle condition, of a vanishing velocity, by studying the interplay of disorder and quasiperiodicity on equal footing. 
A closely related problem has been studied in two-dimensional Dirac semimetal models 
and is pertinent to understand the role of twist disorder in magic-angle graphene experiments~\cite{beechem_rotational_2014, wilson_disorder_2020,uri2020mapping,PhysRevResearch.2.033458,PhysRevResearch.2.043416,gonccalves2021incommensurability}, which have attracted a great deal of attention. 
However, in two dimensions the marginal relevance of disorder removes the AQCP from the problem and does not allow a direct link between the two effects to be exposed.

\begin{figure}
\centering
	\includegraphics[width=\columnwidth]{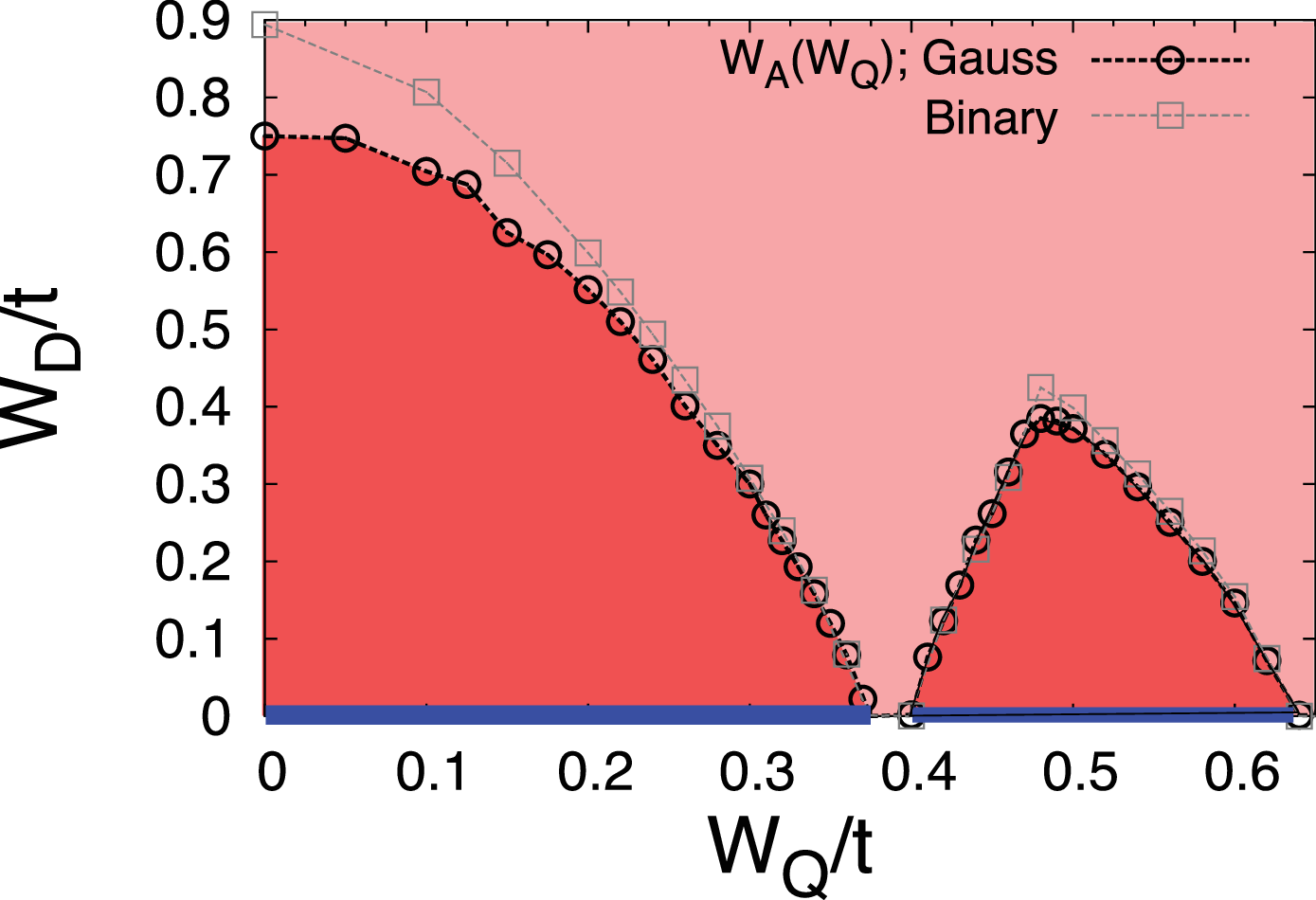}
	\caption{{\bf Phase diagram in disorder ($W_D$) and quasiperiodic ($W_Q$) potential strength at the Weyl node energy ($E=0)$.} 
	Solid blue lines are stable Weyl semimetal phases that terminate at magic-angle transitions (the Weyl semimetal at larger $W_Q$ is an inverted semimetal phase).
	At any non-zero $W_D$ the model is in the diffusive metal phase, dark red marks the semimetal regime $\rho(E)\approx \rho(0) + \rho''(0)E^2/2$ where $\rho(0)$ is nonzero but exponentially small and light red is past the AQCP where $\rho(0)\sim \mathrm{O}(1)$.
	The location of the peak in $\rho''(0)$ as a function of $W_D$ (for fixed $W_Q$) provides an estimate of  the AQCP (and the quasiperiodic transition at $W_D=0$) that we label as $W_A(W_Q)$. We compare two choices of the distribution of the disorder potential $P[V]$ showing their distinction is insignificant near the magic-angle transitions, Gaussian and binary distributions, the latter has been shown to weaken the avoidance for $W_Q=0$ \cite{pixleyUncoveringHiddenQuantum2016}.  This analysis is done on a system size of $L=89$, KPM expansion order $N_C=2^{10}$, and 100 samples. Sufficiently close to the magic angle transitions additional minibands appear that give rise to more structure in $W_A(W_Q)$, see Sec.~\ref{sec:criticalline}; that additional miniband structure is not shown here.
	}
	\label{fig:phase_diagram}
\end{figure}

Through numerical calculations  of the density of states of an inversion-broken 3D Weyl semimetal  using the kernel polynomial method (KPM) we show that the AQCP becomes a line of crossovers that terminate at the MAT, as shown in the phase diagram of Fig.~\ref{fig:phase_diagram}. We study the fate of the analytic properties of the zero energy density of states when disorder is added to the quasiperiodic Weyl semimetal model, demonstrating the interplay of incommensurate induced miniband formation and non-perturbative rare region effects. Last, the critical properties of the magic-angle transition are determined by approaching it along the crossover line of avoided transitions, which allows us to provide an accurate estimate of the power law nature of the vanishing Weyl velocity.

The remainder of the paper is organized as follows: The model and the method used are introduced in Sec.~\ref{sec:model}, we determine the phase diagram of the model in Sec.~\ref{sec:phasediagram}, and the critical properties along the line of cross overs in Sec.~\ref{sec:criticalline}. Finally in Sec.~\ref{sec:discussion} we discuss the implications of our results, its detection using ultracold atoms, and conclude.

\section{Model and Method}
\label{sec:model}

To investigate the interplay of the effects of disorder (D) and quasiperiodicity (Q) on Weyl semimetals we add two separate potentials to a lattice model of an inversion symmetry broken Weyl semimetal given by
\begin{eqnarray}
H &=& \sum_{{\bf r},{\mu}}\left( i t_{\mu}\psi_{{\bf r}}^{\dag}\sigma_{\mu} \psi_{{\bf r}+\hat{\mu}} 
+ \mathrm{H.c}\right)+\sum_{{\bf r}}\psi_{{\bf r}}^{\dag}V(\br)\psi_{{\bf r}}
\label{eqn:weylTR}
\end{eqnarray}
where ${\mu}={x},{y},{z}$, the potential is a sum of two separate contributions from randomness (that we denote with a $D$ for disorder) and quasiperiodicity (denoted with a $Q$)
\begin{equation}
V(\br) = V_D(\br) + V_{Q}(\br),
\end{equation}
which we parameterize below.
The model lives on the simple cubic lattice of linear size $L$ and we average over twisted boundary conditions to reduce finite size effects. The hopping is then given by $t_{\mu}=te^{i \theta_{\mu}/L}/2$ where  the twist in the $\mu$ direction $\theta_{\mu}$ is randomly sampled between $0$ and $2\pi$.
In the absence of the potentials the band structure is given by 
\begin{equation}
    E_0(\bk) = \pm t \sqrt{\sum_{\mu=x,y,z}\sin^2(k_{\mu}+\theta_{\mu}/L)}
\end{equation}
with 8 Weyl cones labeled by ${\bf K}_W$ at the time-reversal invariant momenta (for no twist) in the Brillioun zone. Near each Weyl point $\bK_W$ the dispersion is given by 
\begin{equation}
    E_0(\bk) \approx \pm v_0(\bK_W) |\bk-\bK_W|
\end{equation}
with a velocity $v_0(\bK_W)=\pm t$ that depends on the helicity.

The disorder potential $V_D(\br)$ is sampled independently at each site from a probability distribution $P[V]$. 
In the following we consider two different distributions.
To enhance rare region effects we consider a Gaussian distribution with zero mean and standard deviation $W_D$.
To suppress rare regions and enhance the critical scaling properties we also consider a binary distribution where the value of the potential is equally likely to be $\pm W_D$.
In the absence of the quasiperiodic potential  the semimetal phase of this model is unstable to rare region effects that induce a diffusive metal phase at infinitesimal disorer strength.
The resulting perturbative transition is avoided and rounded into a cross over.
By varying the tails of the distribution $P[V]$ we can control the probability to generate rare events, removing the tails as in the binary case quantitatively suppresses (but does not eliminate) rare region effects~\cite{pixleyUncoveringHiddenQuantum2016}.
All results shown are for the case of Gaussian disorder unless otherwise specified. 

The quasiperiodic potential is given by 
\begin{equation}
    V_{Q}(\br)=W_{Q}\sum_{\mu=x,y,z}\cos(Q_L r_{\mu}+\phi_{\mu})
    \label{eqn:QP}
\end{equation}
where the quasiperiodic wavevector is taken as a rational approximant $Q_L = 2\pi F_{n-2}/L$ where the system size is given by the $n$th Fibonacci number $L=F_n$ and in the thermodynamic limit $Q_L \rightarrow Q=2\pi[2/(\sqrt{5}+1)]^2$. The random phases $\phi_{\mu}$ are randomly sampled between $0$ and $2\pi$ as the origin of the quasiperiodic potential is arbitrary. In the absence of the random potential this model has been shown to host several ``magic-angle''  transitions between Weyl semimetal and diffusive metal phases as a function of increasing $W_Q$. Near the magic-angle transition $W_c$, in the semimetallic phase, the velocity of the Weyl cone vanishes as 
\begin{equation}
    v(W_Q) \sim |W_Q-W_c|^{\beta/d},
\end{equation} 
where $\beta\approx 2$ and $d=3$ is the spatial dimension~\cite{Pixley-2018}.  The plane wave Weyl eigenstates delocalize in momentum space and the level statistics become consistent with random matrix theory when one enters the diffusive metallic phase.
In the following manuscript 
we 
use disorder to round out the critical properties of the quasiperiodic induced transition, which allows us to approach the MAT from a new direction.

To characterize the system we numerically compute the density of states (DOS) that is given by 
\begin{equation}
    \rho(E) = \frac{1}{L^3}\sum_i\delta(E-E_i)
\end{equation}
where $E_i$ are the  eigenenergies of $H$. The DOS is computed using the kernel polynomial method (KPM) by expanding it in terms of Chebyshev polynomials to order $N_C$ and evaluating the expansion coefficients with sparse matrix-vector multiplication.
For the system sizes of $L=55,89$ considered here, $N_C$ is the most dominant finite size effect and therefore we try to converge our results with $N_C$.  
The analytic properties of the density of states are investigated via assuming the DOS is always analytic and Taylor expanding
\begin{equation}
    \rho(E) = \rho(0) + \frac{1}{2}\rho''(0)E^2+\dots
\end{equation}
and we directly
compute the second derivative of the DOS with KPM at the Weyl node energy ($E=0$) $\rho''(0)$~\cite{PixleyBR-2016}.
If the DOS becomes nonanalytic then $\rho''(0)\rightarrow \infty$, whereas if the system undergoes a crossover it will remain finite. 

\begin{figure}[t!]
\centering
	\includegraphics[width=0.7\columnwidth,angle=-90]{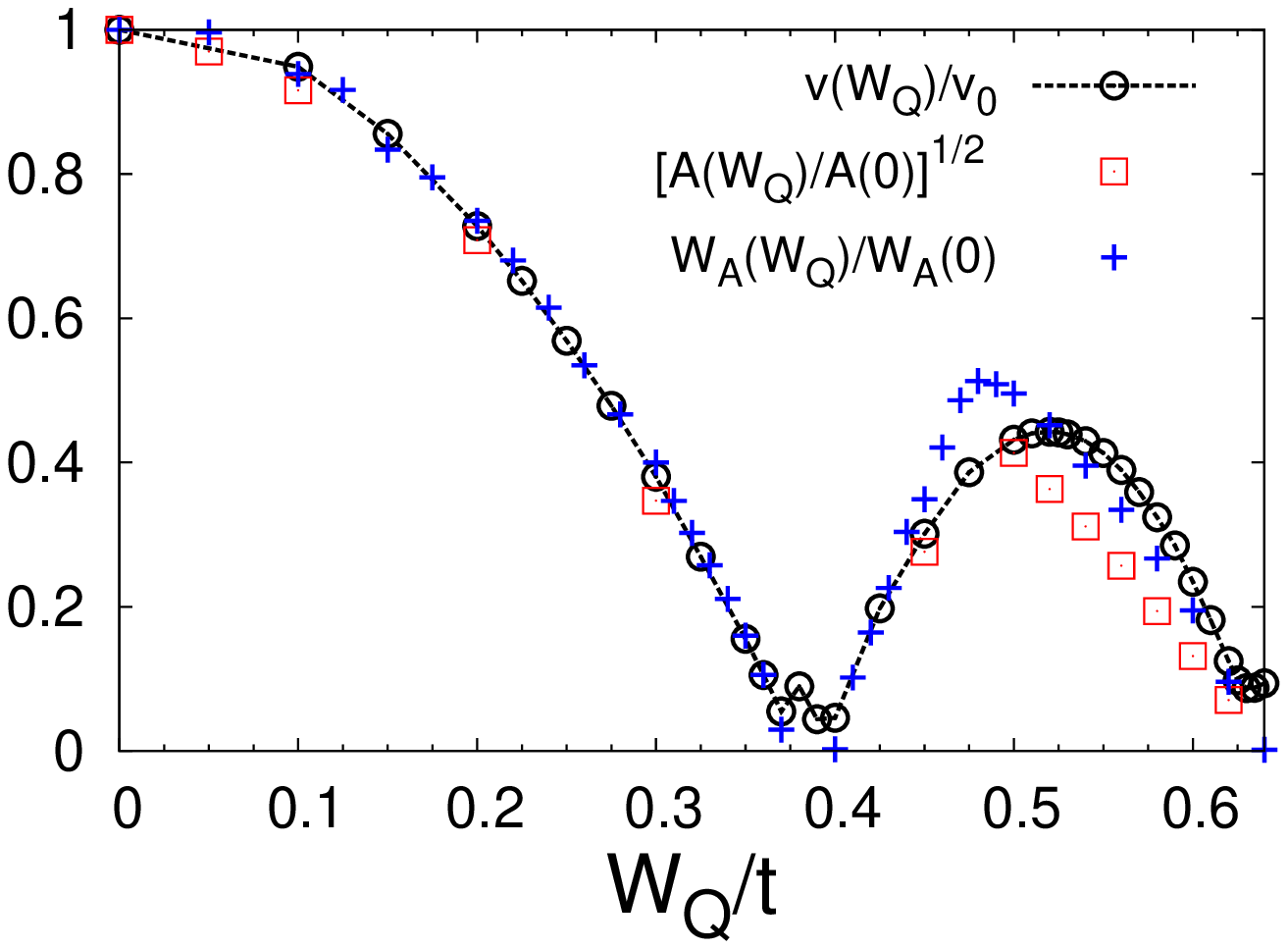}
	\caption{{\bf Renormalized velocity and related scales}:
	 Comparison of the renormalized velocity $v(W_Q)$ with the location of the avoided transition $W_A(W_Q)$ and the dependence of the density of states $A(W_Q)$, see Eq.~\eqref{eqn:rho0}. The velocity $v(W_Q)$ of the renormalized semimetal is obtained from the disorder free limit using $\rho''(0)\propto v^{-3}$ (for system size $L=144$ and $N_C=2^{10}$ from Ref.~\cite{Pixley-2018}), $A(W_Q)$ is extracted  from the rare region dependence of $\log \rho(0) \sim A(W_Q)/(W_D)^2$ (from the data that is converged in $L$ and $N_C$ see Eq.~\eqref{eqn:rho0}), and $W_A(W_Q)$ denotes the line of AQCPs determined from the maximum in $\rho''(0)$ as a function of $W_D$ (obtained from a system size of $L=89$ and KPM expansion order $N_C=2^{10})$.
  Sufficiently close to each MAT the formation of minibands enriches the picture beyond the relations implied by these data; that requires sufficiently large $L$ and or $N_C$ to observe, see Sec.~\ref{sec:criticalline}.  
	}
	\label{fig:velocity}
\end{figure}

For a stable Weyl semimetal phase we have $\rho(0)=0$ and $\rho''(0)=N_W/(2\pi^2v^3)$, where $v$ is the velocity of the Weyl cone and $N_W$ denotes the number of Weyl points in the band structure. Using this relation, an estimate of the velocity in the Weyl semimetal phase  $W_D=0$ is shown in Fig.~\ref{fig:velocity}. Importantly, this also implies that when $v\rightarrow 0$ the DOS becomes non-analytic as $\rho''(0)\rightarrow\infty $ signalling a MAT. Thus, Fig.~\ref{fig:velocity} also 
demonstrates the existence of three 
MATs taking place at $W_{M,1}\approx 0.38t$ into the DM phase, out of the DM phase to a reentrant SM at $W_{M,1}' \approx 0.395 t$ and then a transition back to the DM phase at $W_{M,2}\approx 0.6345t$ for this range of $W_Q$ and $Q/2\pi=[2/(\sqrt{5}+1)]^2$. 
We note that  the reentrant semimetal phase for $W>W_{M,1}'$ occurs by inverting the positive and negative energy bands, which we refer to as an inverted Weyl semimetal phase.

\begin{figure*}[t!]
\centering
	\includegraphics[width=0.45\columnwidth,angle=-90]{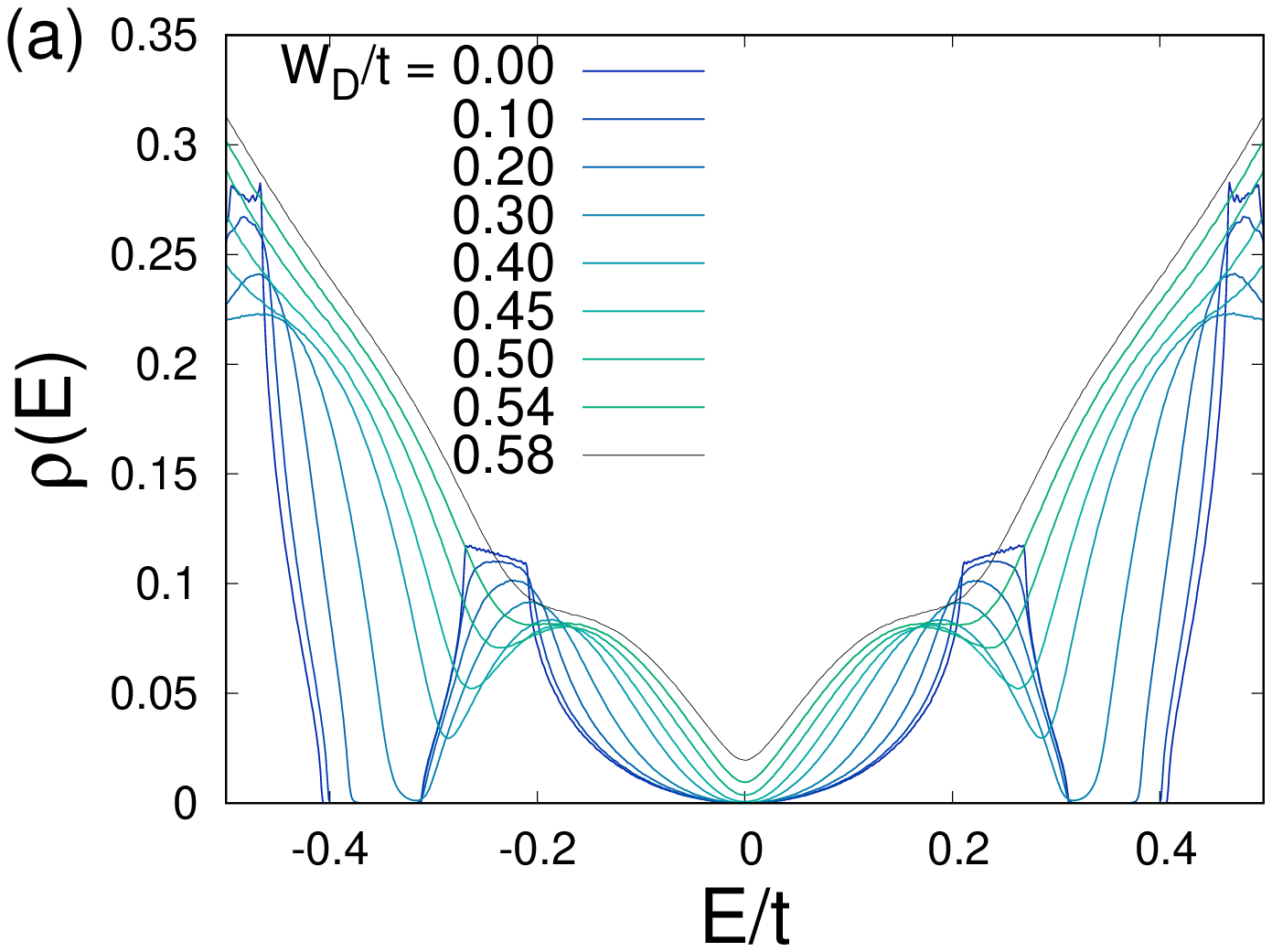}
	\includegraphics[width=0.45\columnwidth,angle=-90]{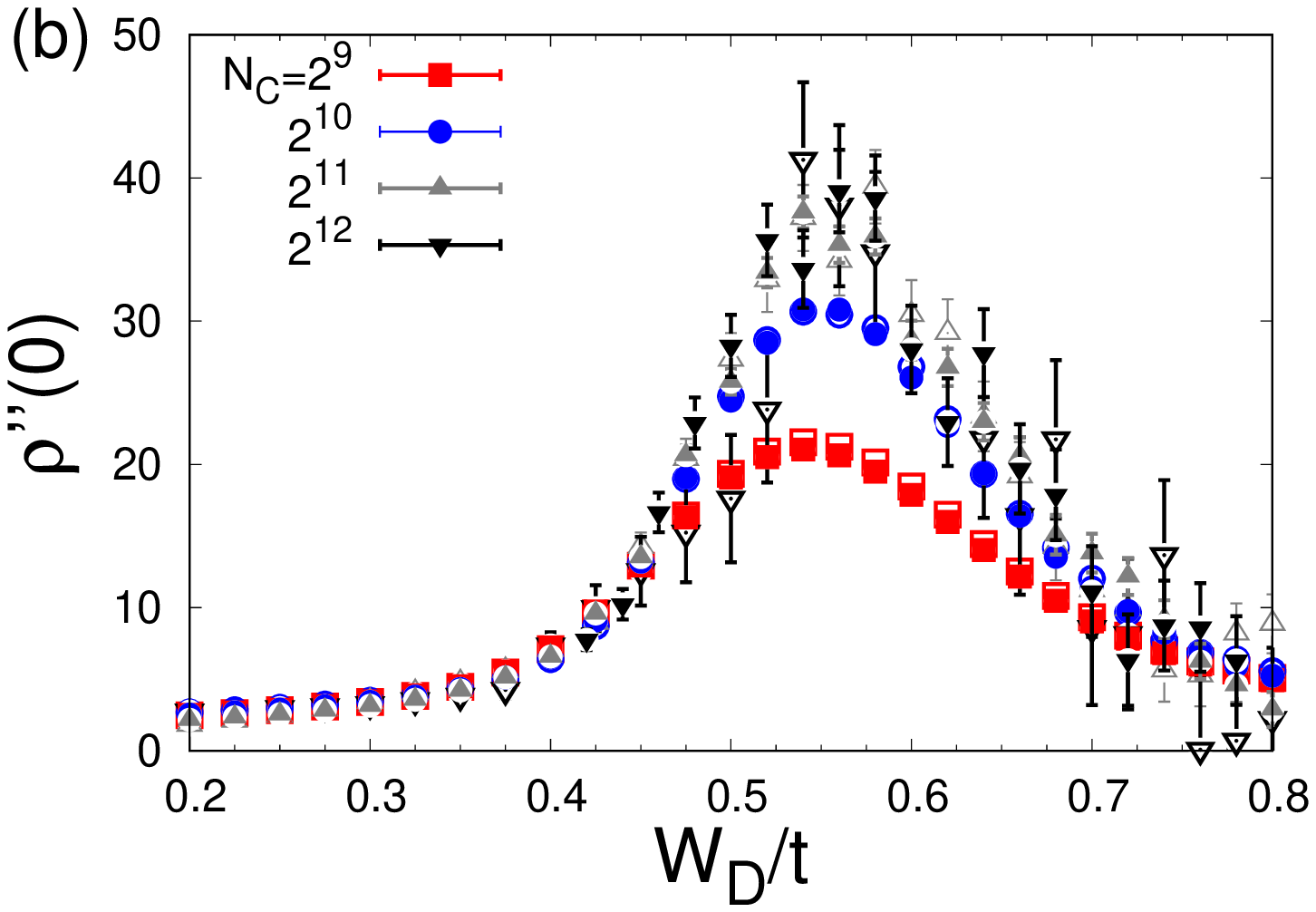}
		\includegraphics[width=0.45\columnwidth,angle=-90]{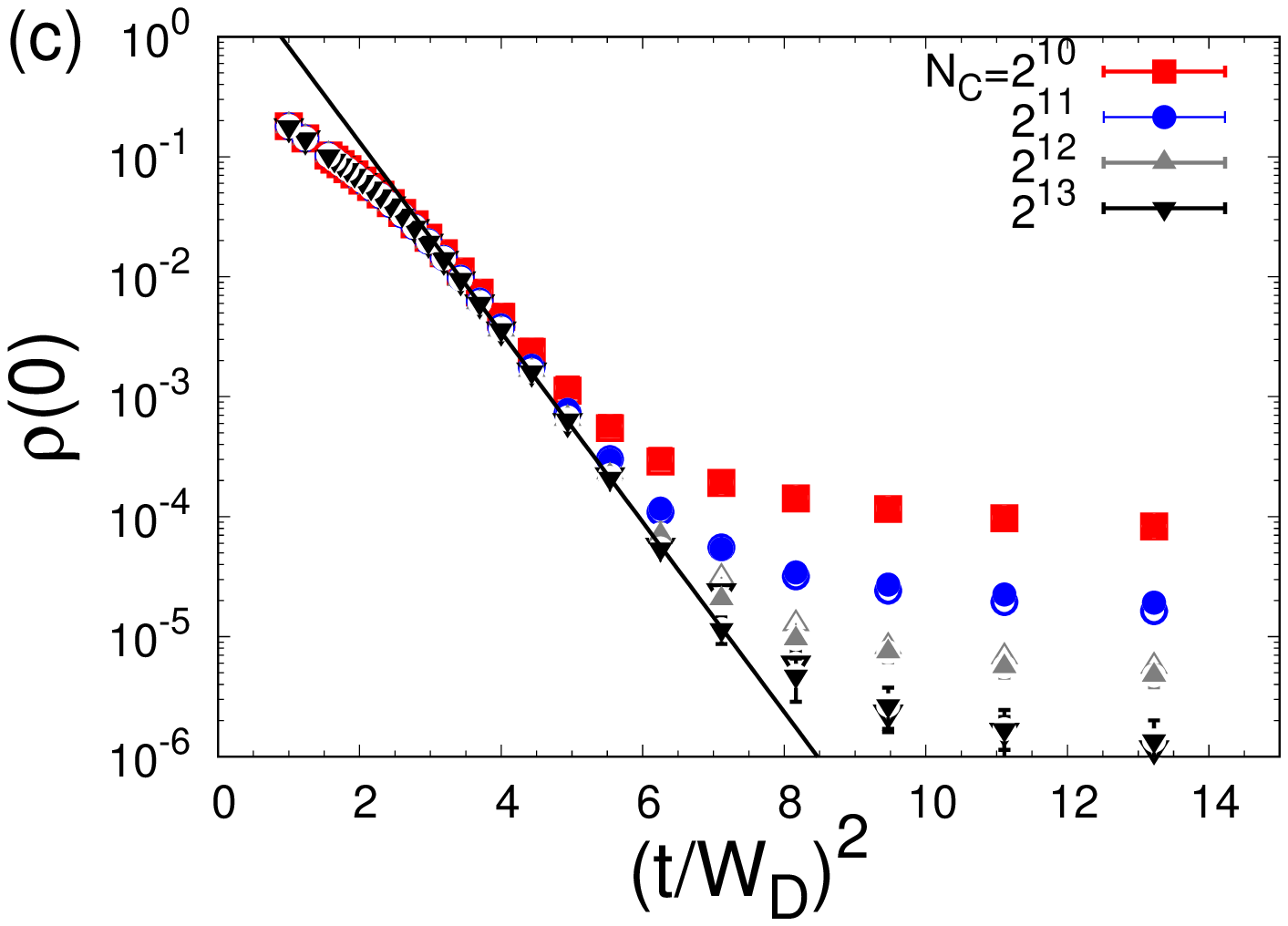}
	\\
\includegraphics[width=0.45\columnwidth,angle=-90]{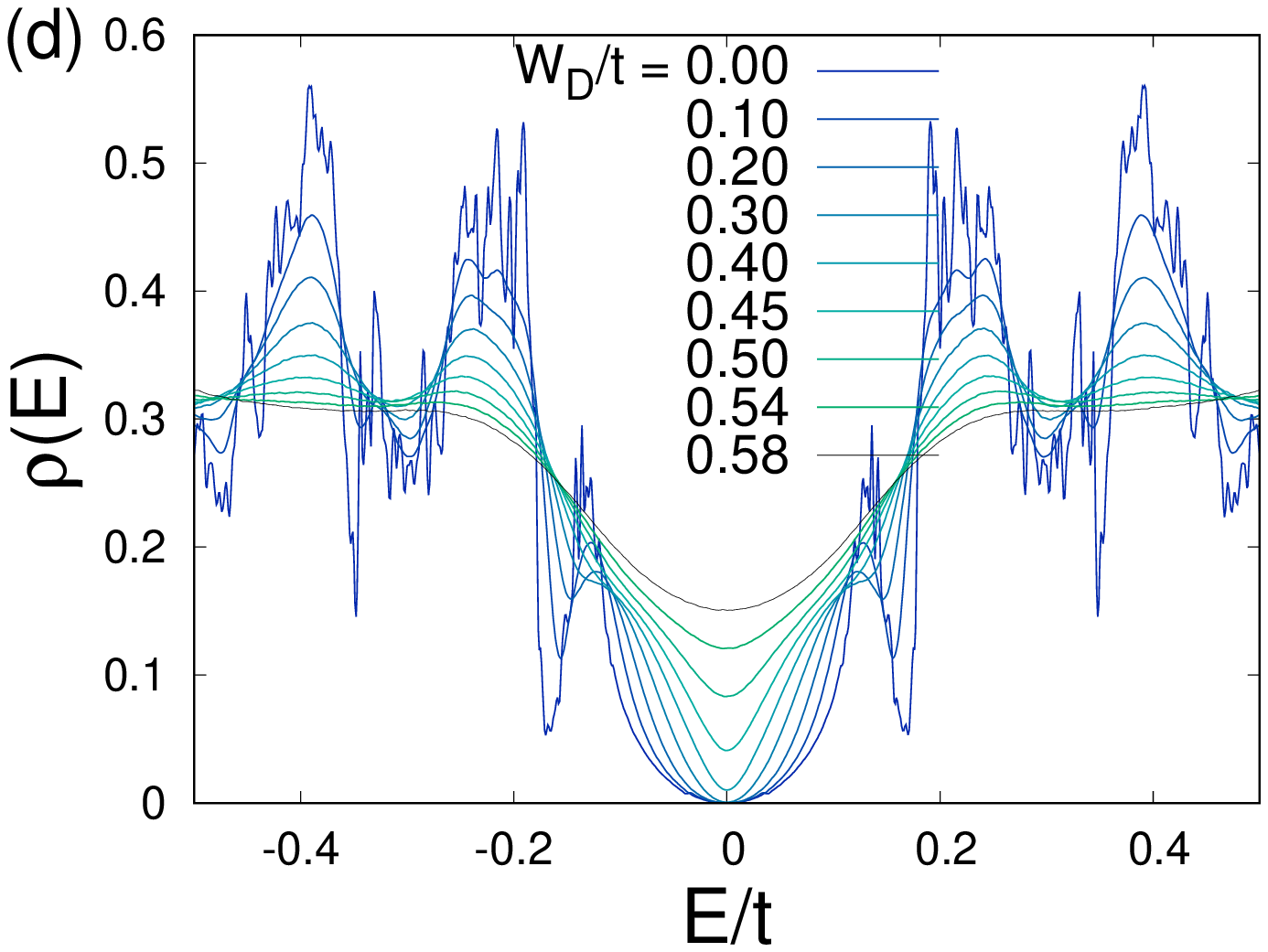}
	\includegraphics[width=0.45\columnwidth,angle=-90]{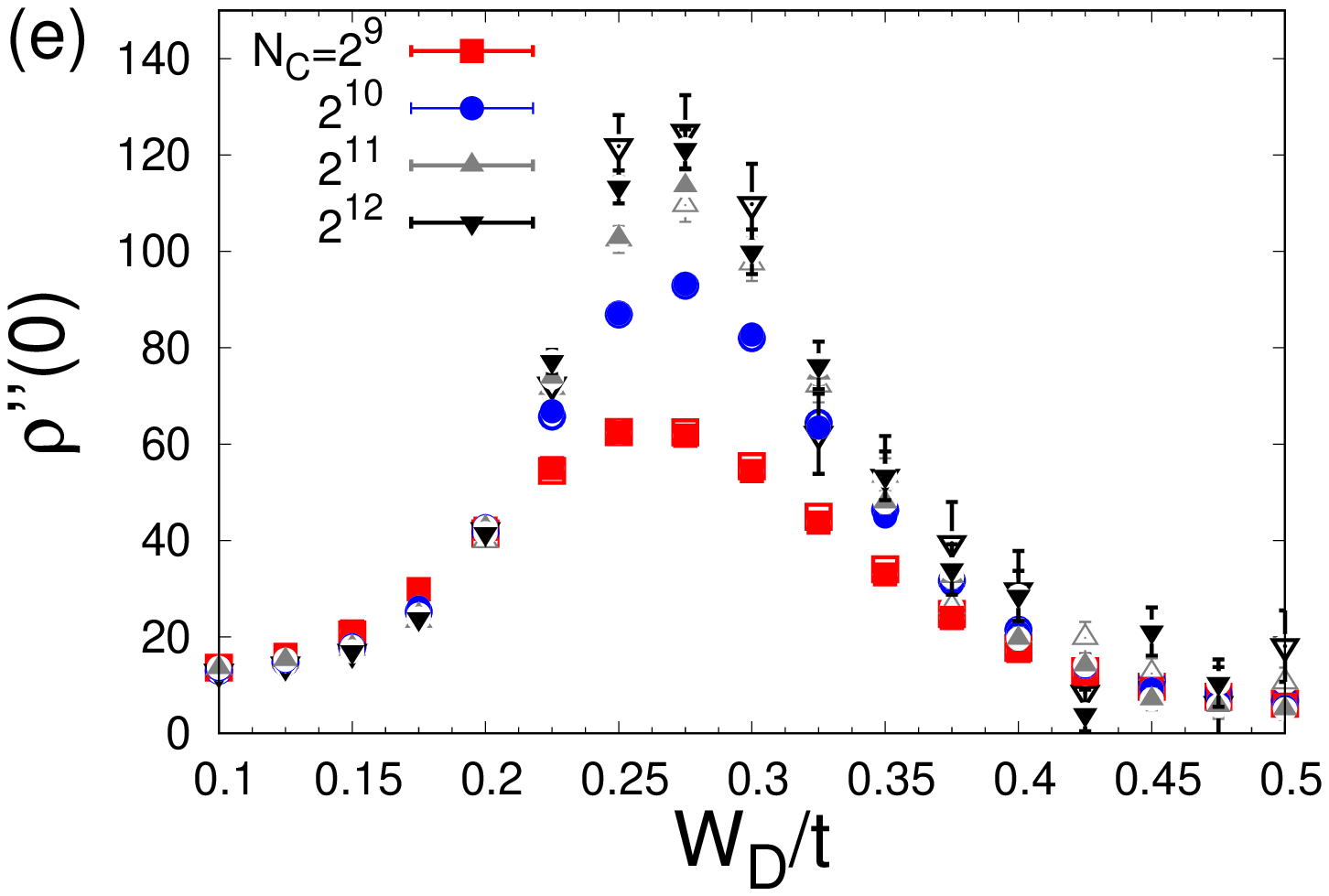}
		\includegraphics[width=0.45\columnwidth,angle=-90]{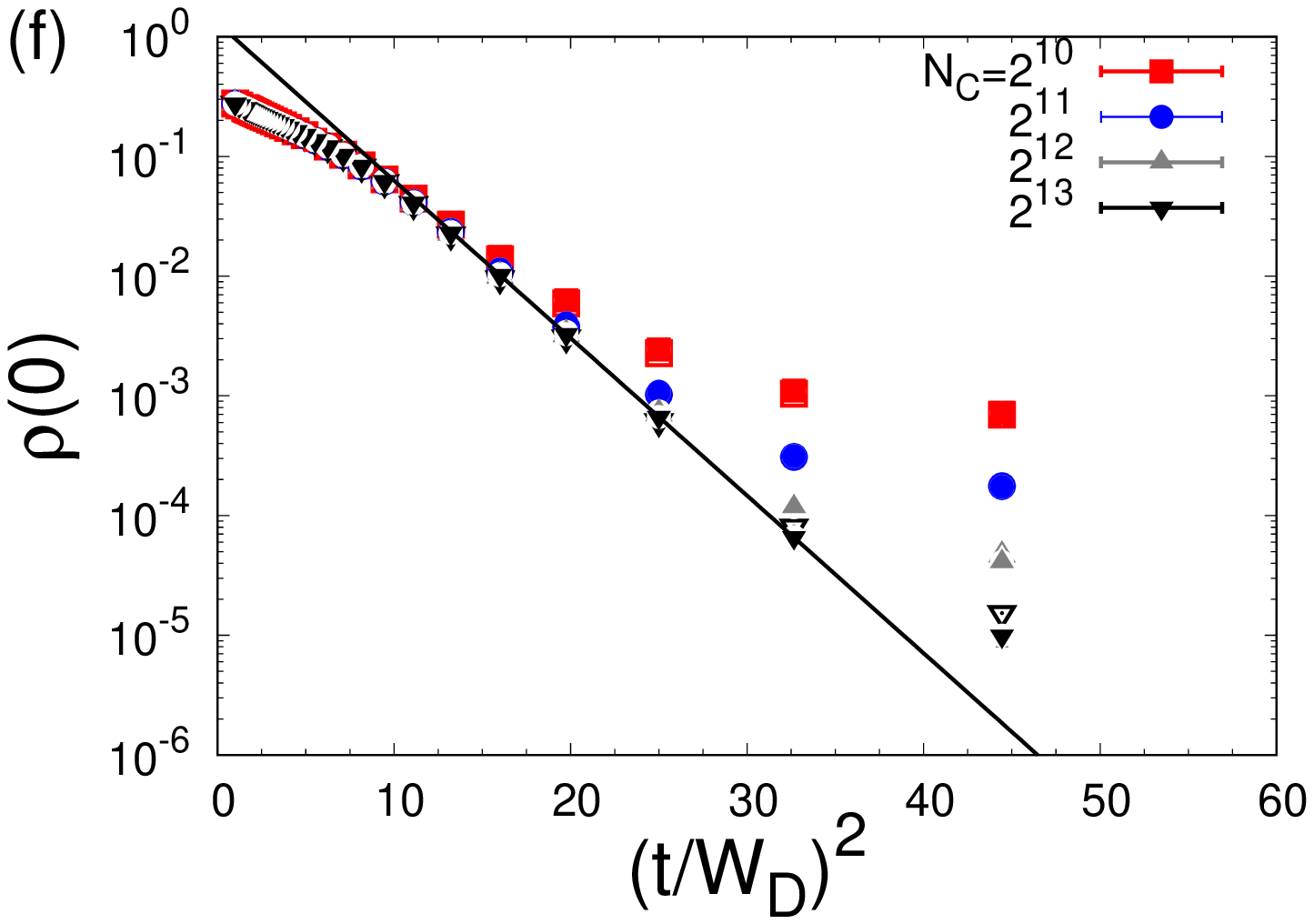}
	\caption{{\bf Evolution of the density of states}: $\rho(E)$ (a), (d) (with $L=89$ and $N_C=2^{13}$); $\rho''(0)$ (b), (e); and $\rho(0)$ (c), (f) as a function of $W_D$ and $W_Q$  for various values of the KPM expansion order $N_C$, $L=55$ (open symbols), and $L=89$ (closed symbols).  
	The values of the quasiperiodic potential are: $W_Q=0.2t$ in the semimetal phase (Top row) and $W_Q=0.55t$ in the inverted semimetal phase (Bottom row). These data are averaged over 1000 samples for $N_C=2^9,2^{10},2^{11}$ and 5000 samples for $N_C=2^{12},2^{13}$. Straight lines in (c) and (f) are fits to the rare region form in Eq.~\eqref{eqn:rho0}.
	}
	\label{fig:rhoE}
\end{figure*}

\section{Phase Diagram}
\label{sec:phasediagram}

As the Weyl semimetal phase is stable in the presence of a quasiperiodic potential we find it most natural to explain the structure of the phase diagram at the Weyl node energy by adding disorder to the quasiperiodic Weyl semimetal model. From this perspective, we can safely use perturbation theory to determine a new low energy effective model in the quasiperiodic renormalized semimetal phase.
We note that for $W'_{M,1}<W_Q<W_{M,2}$ the inverted semimetal phase requires a rather high order in perturbation theory to be described.
For $W_D=0$ we can evaluate the self energy of the single particle Green function by treating $W_Q$ perturbatively~\cite{fuMagicangleSemimetals2020}, similar to what is done to describe twisted bilayer graphene~\cite{bistritzer2011moire}. Focusing on the present case with $Q$ close to $\pi$ we only need to consider internode scattering. This results in a renormalized velocity $v(W_Q)$, with a perturbative expression 
to leading order~\cite{fuMagicangleSemimetals2020}
\begin{equation}
    \frac{v(W_Q)}{v_0} \approx \frac{1-2(2-\cos(Q))\alpha^2}{1+6\alpha^2}
\end{equation}
where 
the dimensionless coupling constant  $\alpha=W/[2 t\sin(Q)]$ and a magic-angle condition  occurs where $v(W_Q=W_\mathrm{MA})=0$. We note that  sufficiently high orders in perturbation theory are required to describe the data in Fig.~\ref{fig:velocity}. Nonetheless,
our numerical results confirm beyond perturbation theory that
the quasiperiodic potential produces a magic-angle transition where the velocity vanishes. At the same time, away from the magic-angle transition the quasiperiodic potential carves out a mini Brillouin zone (mBZ), with an effective band structure on an emergent moir\'e lattice that is qualitatively described by  perturbation theory (see Ref.~\cite{fuMagicangleSemimetals2020} for an explicit construction of the band structure along these lines).

The band gap in the density of states in Fig.~\ref{fig:rhoE}(a) for $W_D=0;W_Q=0.2t$ demonstrates the stability of the Weyl semimetal phase at low energies and the presence of the mBZ .
As we will demonstrate below, our numerical results in a portion of the weakly disordered semimetal phases of the model can thus be interpreted as introducing disorder to a Weyl semimetal that lives on the mBZ with a renoramlized velocity $v(W_Q)$.  At larger quasiperiodic strength, in particular, in the reentrant semimetallic phase with $W_Q\gtrsim0.5t$ this is modified due to the inversion of the bands and  the lack of a true band gap, an example of which is shown for $W_D=0;W_Q=0.55t$ in Fig.~\ref{fig:rhoE}(d). In each case, introducing disorder smoothly fills in these band gaps, pseudogaps, and fine features while rounding out the sharp structure that is due to quasiperiodicity.

To determine the location of the AQCP at finite disorder and quasiperiodic potential strength we evaluate $\rho''(0)$ for fixed $W_Q$ as a function of $W_D$ to determine the location of the peak in $\rho''(0)$ as shown in Fig.~\ref{fig:rhoE}(b,e), which provides an accurate estimate of the AQCP crossover location $W_A(W_Q)$. Importantly, this data is converged in system size $L$ and KPM expansion order $N_C$ and there is no divergence of $\rho''(0)$ upon increasing either $L$ or $N_C$ demonstrating  the  cross over nature of the AQCP. 
Doing this across the parameter regime results in the phase diagram shown in Fig.~\ref{fig:phase_diagram}. 
We note that the phase boundary is obtained for a fixed system size $L=89$ and KPM expansion order $N_C=2^{10}$ and very close to the MAT at small $W_D$ it could be weakly shifted.
Remarkably, the AQCP smoothly connects from the termination of the small diffusive metal phase due to the first magic-angle transition at at $W_Q=W_{M,1}'$ to the 
second magic-angle transition near $W_Q=W_\mathrm{MA,2}\approx 0.63t$.  
Comparing the cross over boundary in Fig.~\ref{fig:phase_diagram} with the estimates of the velocity of the disorder free model $v(W_Q)$ shown in Fig.~\ref{fig:velocity} demonstrates that the line of avoided transitions $W_A(W_Q)$ is simply parameterized by the relation
\begin{equation}
    W_A(W_Q)\propto v(W_Q),
    \label{eqn:WA}
\end{equation}
for $W_Q<W_{M,1}$.
This relation inside the semimetal phase demonstrates that in the low energy limit the only relevant scale left in the problem is the Weyl cone velocity $v(W_Q)$, which we comment on in more detail at the end of this section.

In Fig.~\ref{fig:phase_diagram} we compare the line of AQCPs between Gaussian and binary disorder distributions. The distinction between these two distributions is only significant at sufficiently weak $W_Q$. This can be understood as follows: As we increase the quasiperiodic potential strength (with $W_D=0$) a semimetal miniband forms around zero energy for $W_Q\gtrsim 0.15t$ with a hard gap [e.g. Fig.~\ref{fig:rhoE}(a)] and a new effective mini Brillioun zone of linear size $(\pi-Q)/a$ (where $a=1$ denotes the lattice spacing). As a result, an emergent unit cell develops that goes from  size $a$ to $a_\mathrm{MB}=a \frac{2\pi}{(\pi-Q)}\approx 8.5 a$. 
As the magic-angle transition is approached further, lower energy minibands continue to appear~\cite{fuMagicangleSemimetals2020}.
If we then project the Hamiltonian onto the lowest energy miniband, that is separated from the rest of the states by a hard gap via the projection operator $\hat P_\mathrm{MB}=\sum_{E_n \in MB}|E_n\rangle\langle E_n|$ (that sums over energy eigenstates with energies within the miniband), we can then compute Wannier states on the lowest energy miniband. Thanks to the hard gaps and no topology in the band structure
these   are exponentially localized to Wannier centers~\cite{RevModPhys.84.1419} labeled by $\bR$ on the moir\'e lattice with the Wannier functions $W_\bR(\br)$. Applying this unitary operation plus a projection onto the lowest energy miniband maps the disorder potential in the Hamiltonian to 
$
    \sum_{\br}V_D(\br)\psi_{\br}^{\dag}\psi_{\br}\rightarrow
    \sum_{\bR} \tilde V_D(\bR)\psi_{\bR}^{\dag}\psi_{\bR}
$
where $\bR$ labels the Wannier centers and 
\begin{equation}
    \tilde V_D(\bR)=\sum_{r\in a_\mathrm{MB}}V(\br)|W_\bR(\br)|^2
\end{equation}
is the coarse grained random potential on the scale of the moir\'e unit cell.
As a result, the sharp distinction between the Gaussian potential that has large local fluctuations and the binary distribution that does not is lost after coarse graining over this larger unit cell. This conclusion  is consistent with the lack of a distinction between Gaussian and binary disorder in the vicinity each magic-angle transition.

\begin{figure}
\centering
	\includegraphics[width=0.7\columnwidth,angle=-90]{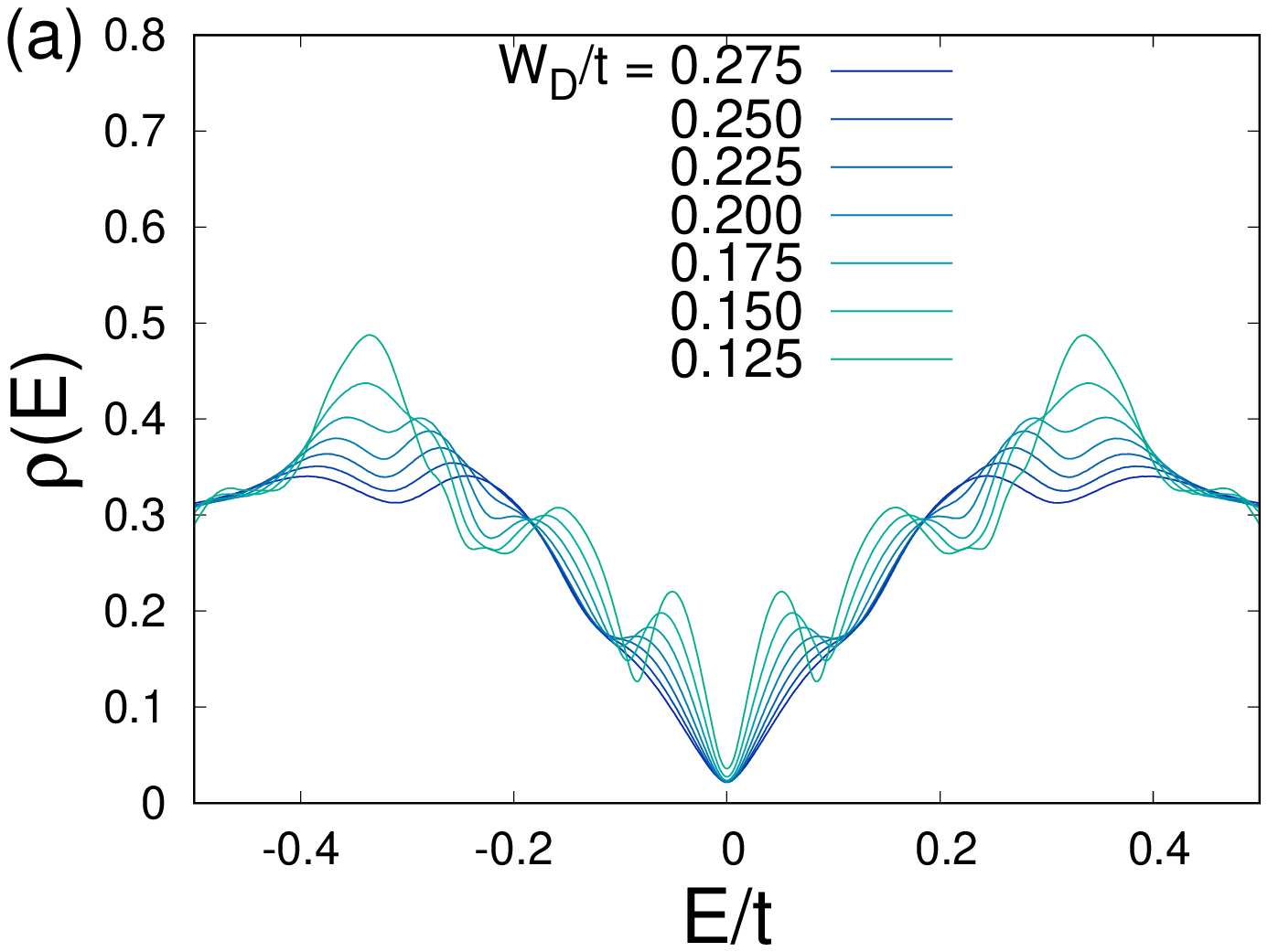}
	\\
	\includegraphics[width=0.7\columnwidth,angle=-90]{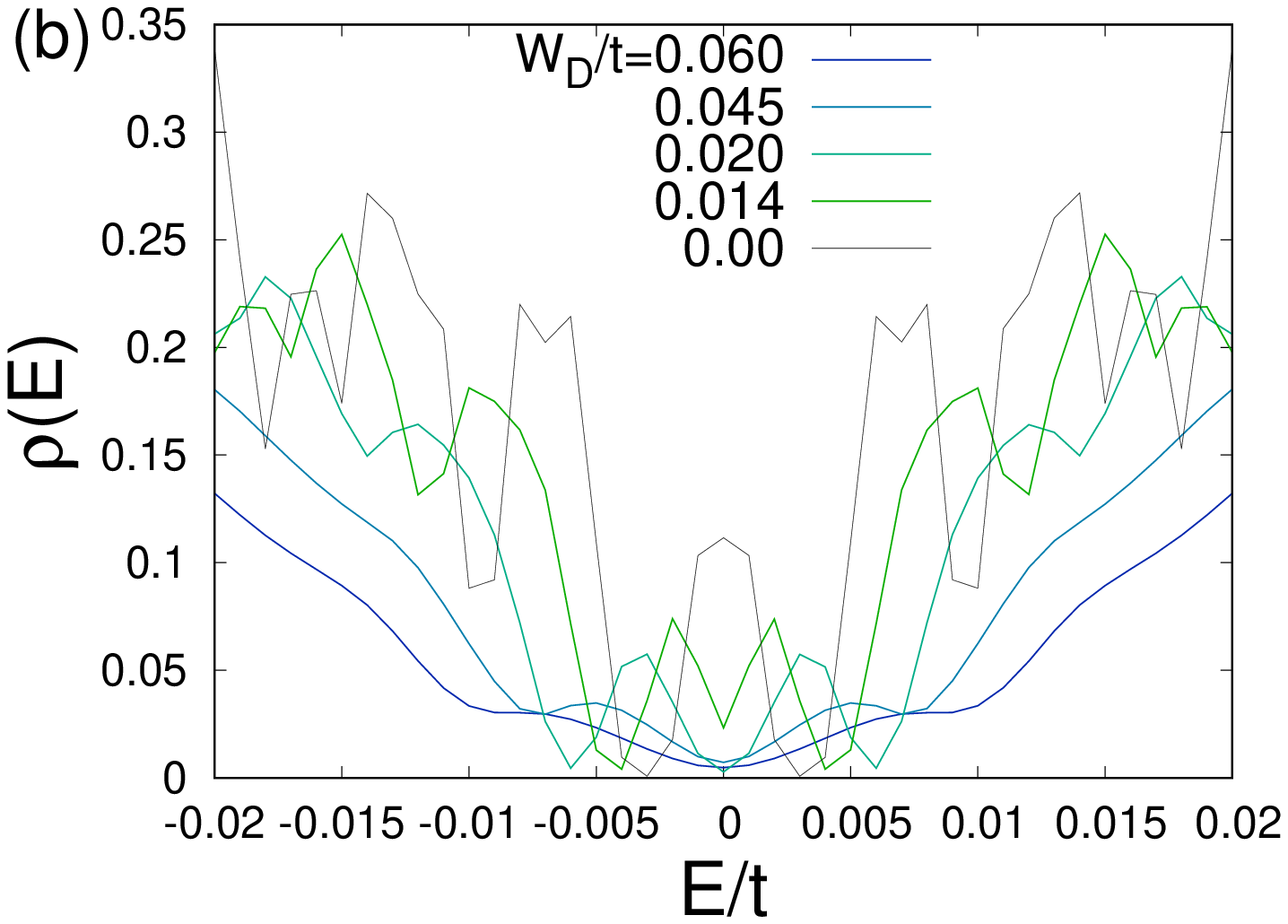}
	\caption{{\bf Density of states  along the line of AQCPs that is defined by $W_{A}(W_Q)$}. We show $\rho(E)$ for $L=89$ at larger $W_D=W_A(W_Q)$   in  (a) (with $N_C=2^{13}$) and closer to the transition at a much lower energy scale in (b) (with $N_C=2^{14}$) as $W_A \rightarrow 0$,  (recall $W_A(W_Q)$ is shown in the phase diagram in Fig.~\ref{fig:phase_diagram}). The data is shown  starting from $W_Q=0.55t$ and terminating at the magic-angle transition at $W_Q^c= 0.6345t$ for $W_D=0$. The finite but low energy dependence along the line of AQCPs is consistent with the expected behaviour in Eq.~\eqref{eqn:rhoEE}. However, as we go from panel (a) to (b) we can see that a consequence of the second miniband opening up for $W_Q\approx 0.62 t$ realizes a dramatically renormalized $\rho''(0)$ seen through the shape near zero energy until we hit the MAT and a finite density of states is generated ($W_D=0$).
 }
	\label{fig:criticalrhoE}
\end{figure}

Now turning on finite disorder strength at a fixed value of $W_Q$ (that remains in the semimetal phase), we expect that a non-zero density of states will be induced at the Weyl node energy due to rare-region effects. As shown in Fig.~\ref{fig:rhoE}(c) and (f) we find that the DOS is converged in $N_C$ and goes like 
\begin{equation}
    \log \rho(0) \sim -\frac{A(W_Q)}{W_D^2}
    \label{eqn:rho0}
\end{equation}
(for each value   $W_Q $ that is in the semimetal phase of the model for $W_D=0$).
From fits to this rare region form as shown in Figs.~\ref{fig:rhoE}(c) and (f) we extract $A(W_Q)$.
At weak disorder and quasiperiodicity strength we find good agreement with the identification 
\begin{equation}
    A(W_Q)\propto v(W_Q)^2
    \label{eqn:A}
\end{equation}
as demonstrated in Fig.~\ref{fig:velocity} by comparing to the numerically estimated value of $v(W_Q)$ from $\rho''(0)$.
However, we do find a distinction at larger quasiperiodic strengths $W_Q \gtrsim 0.5t$ where the Weyl semimetal miniband is no longer isolated from the rest of the states by a hard gap such as in Fig.~\ref{fig:rhoE}(d) (this minigap closure occurs near $W\approx 0.5t$ that is not shown), which alters the  shape of the cut off function and energy dramatically. Correspondingly the prefactor in the DOS $A(W_Q)$ doesn't simply follow $v(W_Q)$ in this regime ($W_Q\gtrsim 0.5t$).

\begin{figure*}[t!]
\centering	\includegraphics[width=0.47\columnwidth,angle=-90]{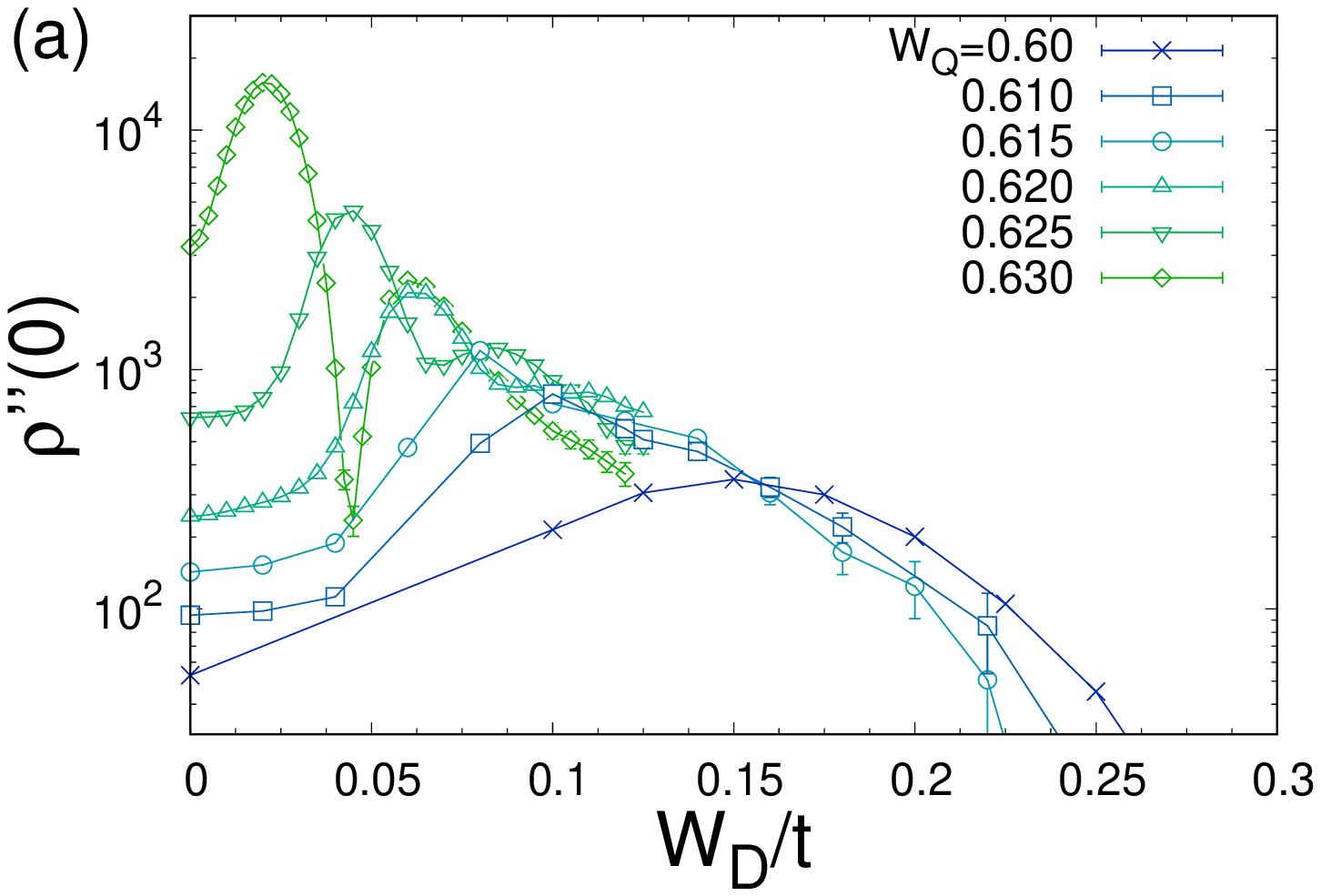}	
\includegraphics[width=0.47\columnwidth,angle=-90]{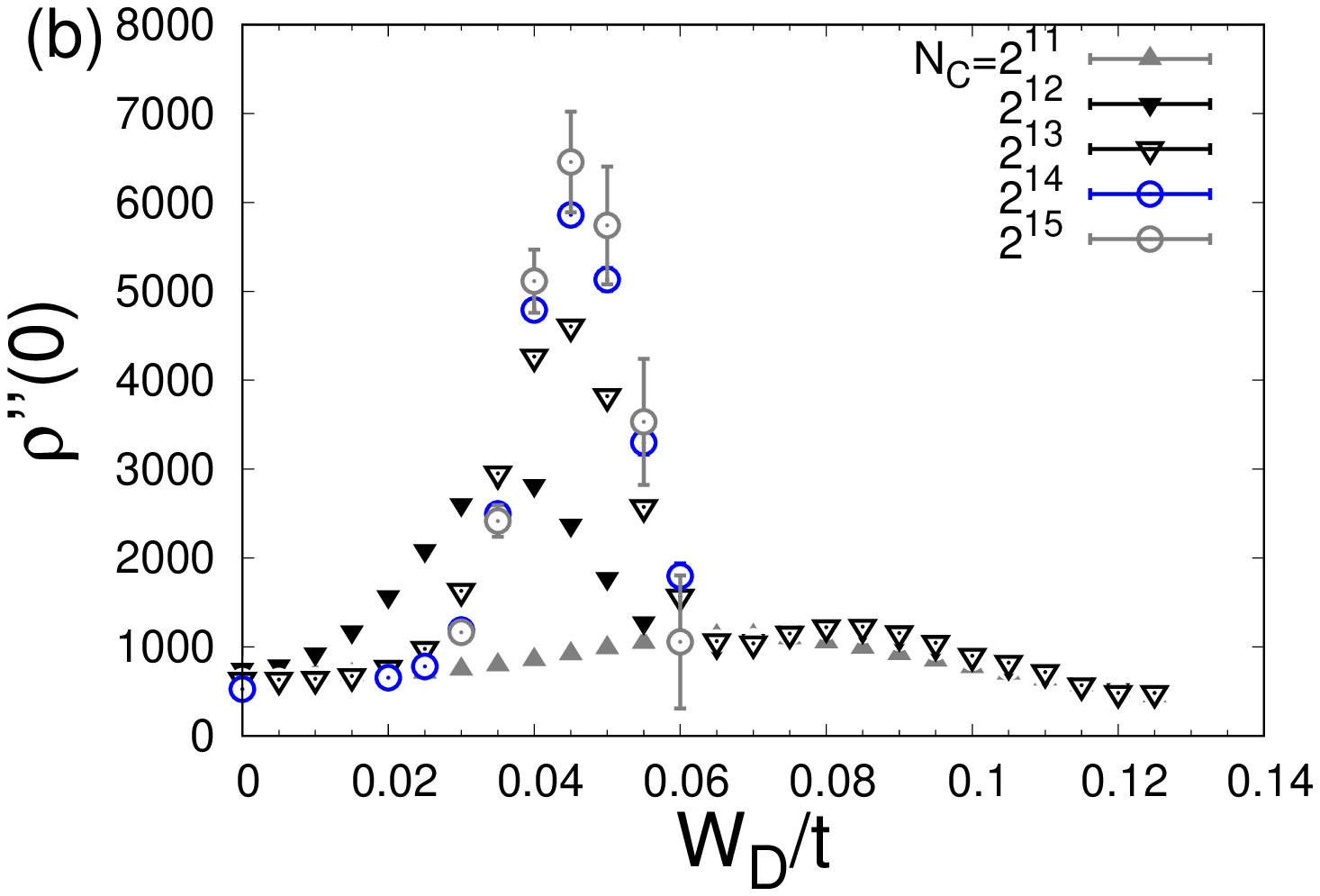}
	\includegraphics[width=0.47\columnwidth,angle=-90]{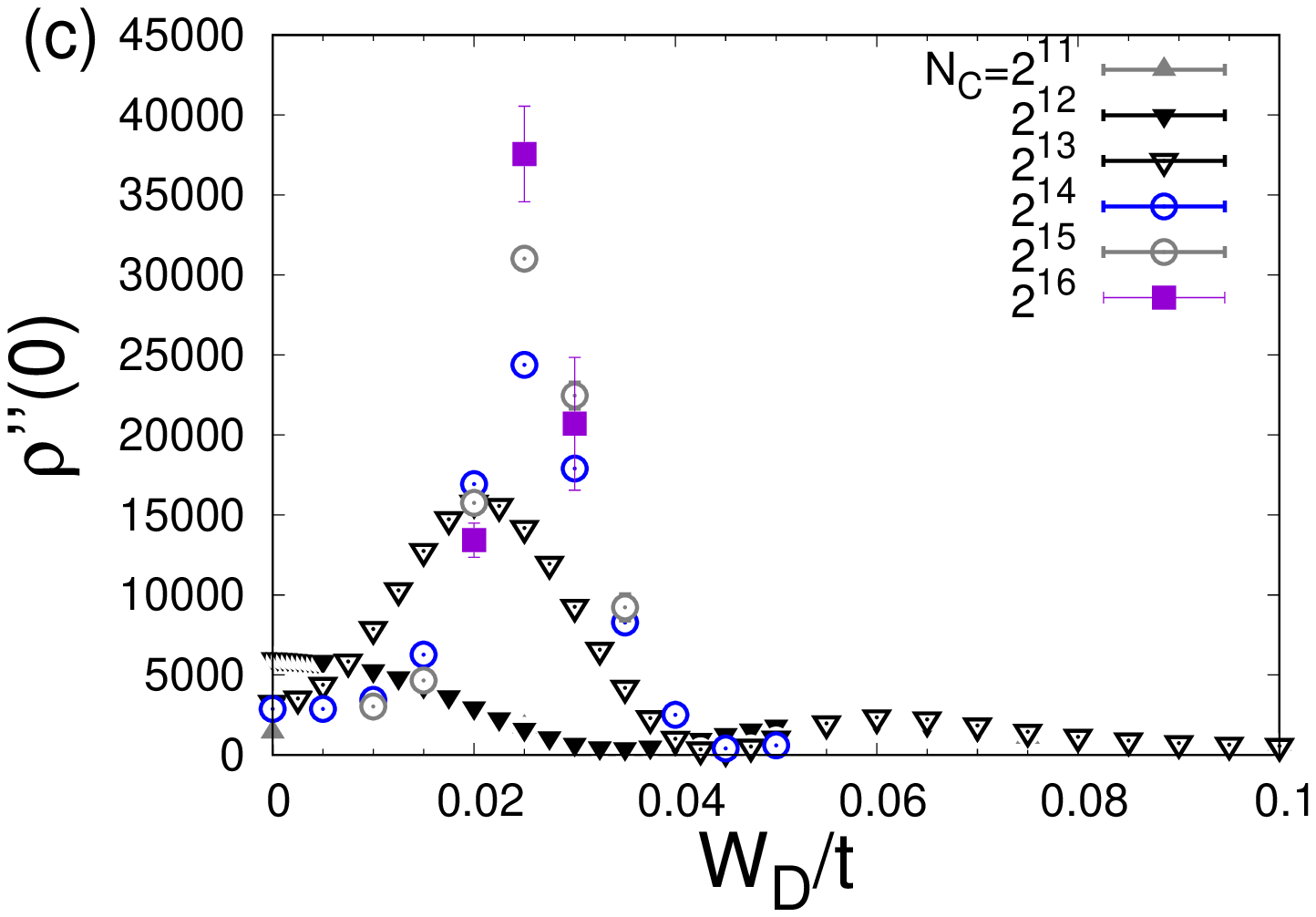}
	\caption{{\bf Evolution of the AQCP on approach to the magic-angle transition   along $W_A(W_Q)$.} (a) Focusing on $\rho''(0)$ on approach to the transition for $N_C=2^{13}$ we see that the original peak that we associate to the AQCP splits off leaving behind a much weaker second peak at larger $W_D$ that can be associated with an approximate AQCP  due to the parts of the band outside of the second miniband. (b) We study the evolution of the peaks for $W_Q=0.625t$ as a function of the expansion order demonstrating a converged AQCP peak at this system size ($L=89$). However, for $W_Q=0.63t$ as we get closer to the MAT we are unable to converge the peak, see also Fig.~\ref{fig:criticalrho2}(a).}
	\label{fig:criticalrho1}
\end{figure*}

The above data and results in Eqs.~\eqref{eqn:WA} and \eqref{eqn:A} suggests that in the semimetal phase there is only one relevant scale $v(W_Q)$ and described in the following picture.
For a Weyl semimetal in the presence of disorder, the low-energy continuum model $H = v \mathbf k \cdot \sigma + V(\mathbf x)$ has one dimensionless parameter that controls the physics: $\alpha_D = W_D/(v/a)$ for disorder strength $W_D$, velocity $v$, and cutoff (lattice) scale $a$.
As the length scale is not varied in this problem, only $v = v(W_Q)$ in the low-energy, suggesting $W_A \sim v(W_Q)$ as we find.
In a similar manner, the density of states should be exponentially suppressed by $\log \rho(0) \sim -1/\alpha_D^2 \sim - v(W_Q)^2/W_D^2$.
This simple single parameter which controls the low-energy theory is consistent with all of the data, allowing us to even discover properties of the low-energy Hamiltonian.
It further lends credence to the statement that what we are witnessing is physics occurring due to the Weyl point itself and not other structure imposed by say the myriad of gaps opened by either subtle tight-binding model effects nor the fine structure emergent at higher energies due to quasiperiodicity.

\section{Approaching the MAT along the line of AQCPs}
\label{sec:criticalline}

Having identified the cross-over boundaries marked by the line of AQCPs, defined by  the disorder strength $W_D=W_{A}(W_Q)$ that terminates at the magic-angles i.e. $W_{A}(W_\mathrm{MA})=0$, we now study the critical properties of the magic-angle transition. 
The inclusion of disorder allows us to approach the magic-angle transition from the line of avoided transitions, which are effectively parameterizing the path through parameter space of  maximal correlation length (as a function of $W_D$) for each value of $W_Q$. For concreteness we focus on the second magic-angle that occurs for $W_\mathrm{MA,2}/t\approx 0.63$ and approach it from below.

\begin{figure}
\centering
	\includegraphics[width=0.7\columnwidth,angle=-90]{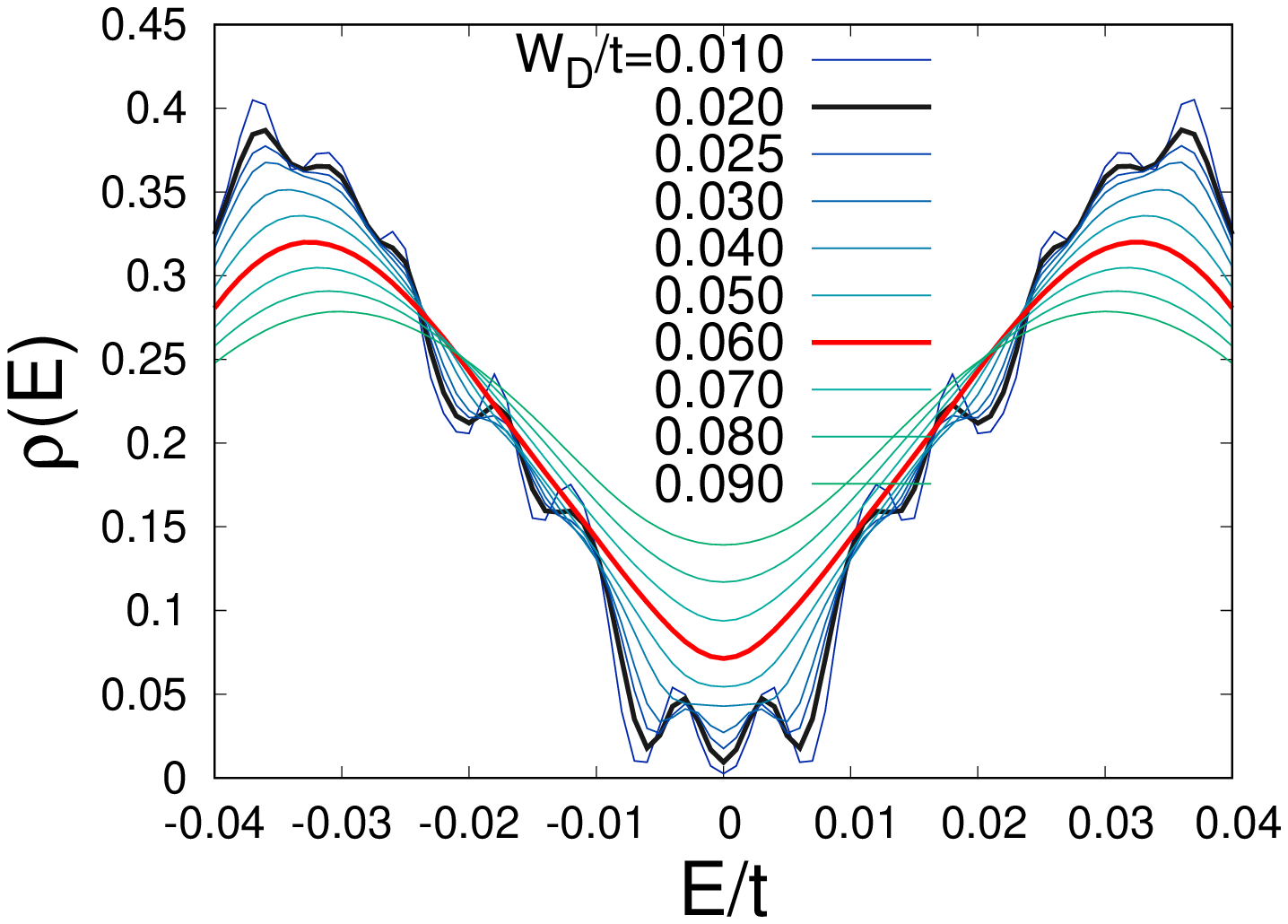}
	\caption{{\bf Demonstration of AQCPs in the first and second miniband for $W_Q/t=0.63$}. We show the density of states $\rho(E)$ as a function of energy $E$ for $L=89$ and $N_C=2^{13}$ at the first maximum in $\rho''(0)$ in black depicting the scaling $\rho(E)\sim |E|$ in the second miniband. The location of the second weaker peak in $\rho''(0)$ is marked in red that depicts the scaling $\rho(E)\sim |E|$ in the first miniband. For reference, see Fig.~\ref{fig:criticalrho1}(a) for the structure of $\rho''(0)$. At each peak in $\rho''(0)$ we find the low energy dependence follows the AQCP form in Eq.~\eqref{eqn:rhoEE} allowing us to identify the signatures of two AQCPs as a function of $W_D$ sufficiently close to the MAT. 
 }
	\label{fig:rhoE-2AQCPs}
\end{figure}

Fig.~\ref{fig:criticalrhoE}(a) and (b) shows the energy dependence of the density of states along the line of avoided transitions terminating at $W=W_\mathrm{MA,2}$. At each avoided critical point the density of states develops the scaling at finite energy [$|E|>|E^*|$, where $E^*$ marks the cross over energy set by the finite value of $\rho(0)$]
\begin{equation}
    \rho(|E|>|E^*|)  \sim |E|
    \label{eqn:rhoEE}
\end{equation}
where this power law is consistent with the one-loop renormalization group results that produces a dynamic exponent $z=3/2$~\cite{Goswami-2011}. 
We note that this energy dependence is seen at each avoided transition (i.e. maximum in $\rho''(0)$ versus $W_D$) as also seen in Fig.~\ref{fig:rhoE}(a) and (d).
However, at the AQCP and sufficiently low energy, the density of states is non-zero $\rho(0)>0$ (as exemplified in Figs.~\ref{fig:rhoE}(c,f) and Eq.~\eqref{eqn:rho0}), which rounds out this power law.  At the MAT on the other hand (with $W_D=0$), $\rho(0)$ is non-zero as seen in Fig.~\ref{fig:criticalrhoE}(b). One can clearly see that the slope of the linear part of the density of states that follows Eq.~\eqref{eqn:rhoEE} is monotonically increasing as we approach the MAT, which is directly reflected in the behavior of $\rho(E)$ and $\rho''(0)$ at the lowest energies, that we now come to.

First, we recall that $\rho''(0)$  diverges at the magic-angle transition~\cite{Pixley-2018}, by approaching this singular behavior from finite disorder strength, it allows us to approach the magic-angle transition in a unique way to probe the critical scaling properties. We first focus on $\rho''(0)$ along $W_A(W_Q)$ at large enough KPM expansion order to resolve the second miniband opening near $W_Q\approx 0.62 t$ ($W_D=0$) in Fig.~\ref{fig:criticalrho1}(a). Before the second miniband opens ($W_Q<0.62 t$) we see a single clear maximum in $\rho''(0)$. In contrast after the second miniband has opened ($W_Q>0.62 t$), we see the AQCP becomes significantly sharper, leaving behind a second peak. While we are able to converge both of these peaks in $N_C$ for $W_Q=0.625t$ as shown Fig.~\ref{fig:criticalrho1}(b) this is not possible as we get closer to the MAT. As an example, we show $W_Q=0.63t$ in Fig.~\ref{fig:criticalrho1}(c), while we are able to converge the weaker peak at larger $W_D$ we cannot converge the sharper weak at weaker disorder strength. 

In order to associate these two peaks with AQCPs in the first and second miniband we show the evolution of $\rho(E)$ at fixed $W_Q=0.63t$ as a function of $W_D$ in Fig.~\ref{fig:rhoE-2AQCPs}. Importantly, we find that at each peak in $\rho''(0)$, $\rho(E)$ follows the expected AQCP ``scaling'' form in Eq.~\eqref{eqn:rhoEE}. As the value of $\rho''(0)$ is significantly larger and not fully converged in $N_C$ for the second miniband we now turn to how this is beginning to diverge as we come to the MAT.

\begin{figure}
\centering
\includegraphics[width=0.7\columnwidth,angle=-90]{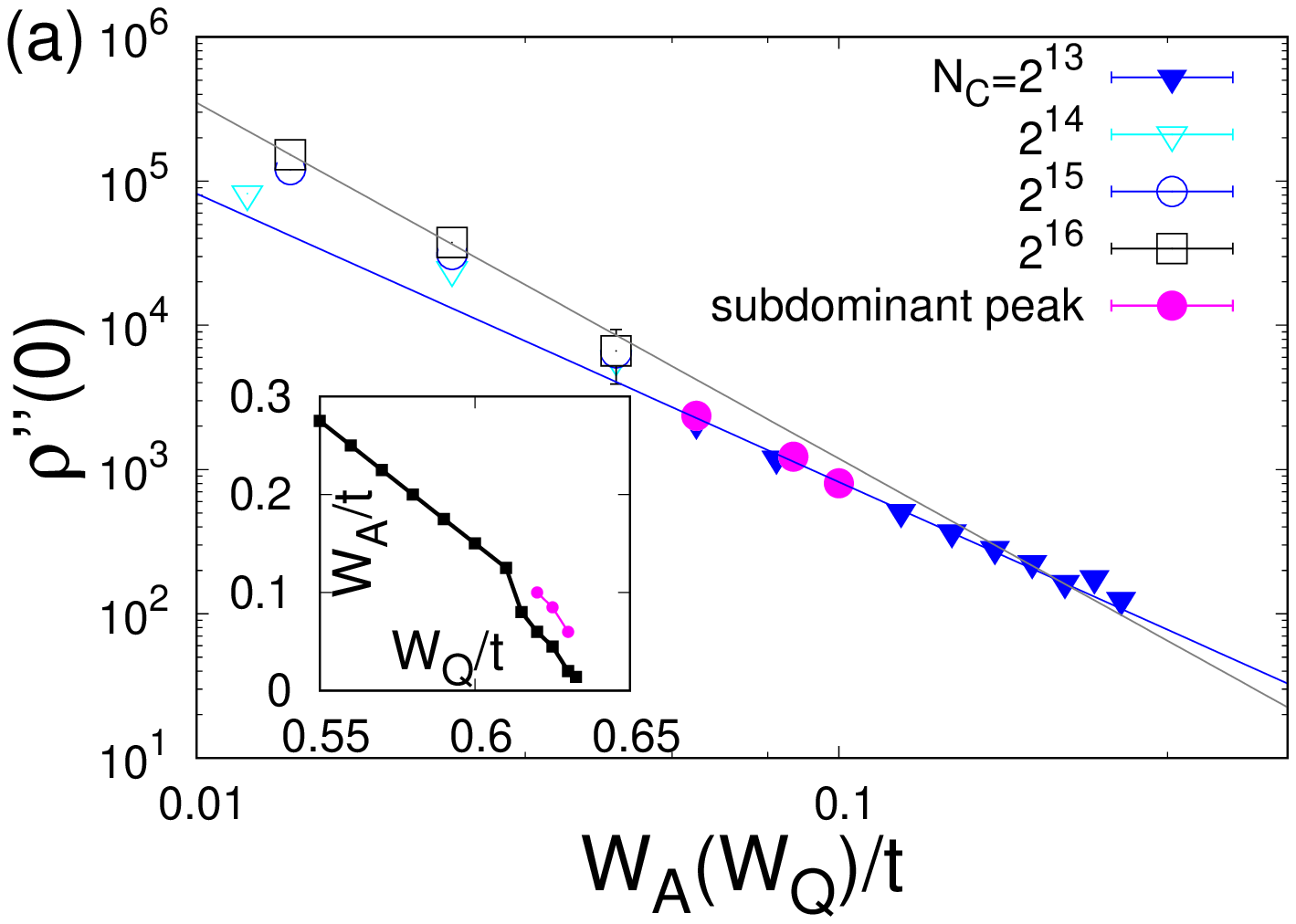}
\\
\includegraphics[width=0.7\columnwidth,angle=-90]{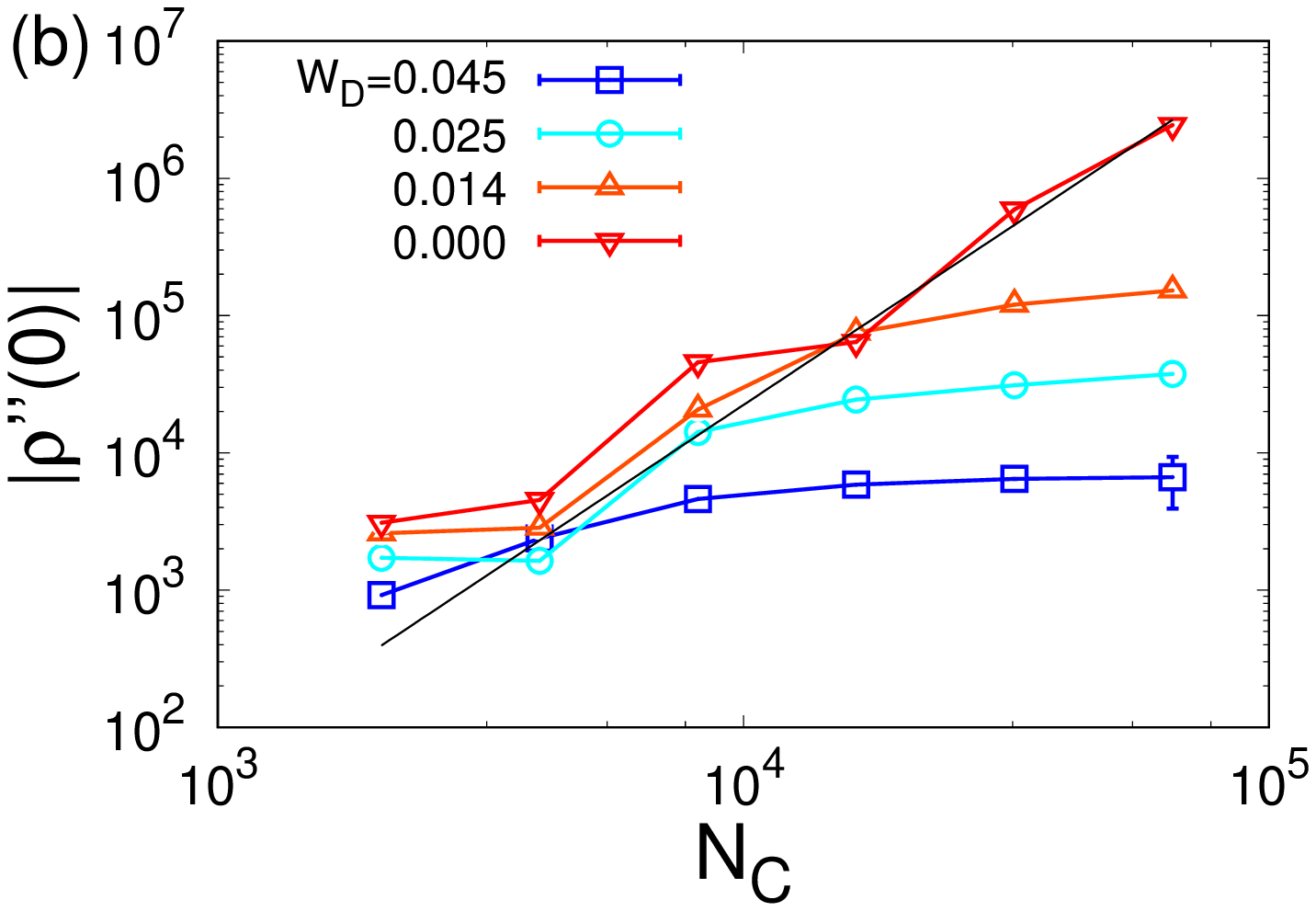}
	\caption{{\bf Divergence of $
 \rho''(0)$ on approach to the MAT along the line of AQCPs}. 
 (a) The solid data points represent peaks in $\rho''(0)$ that are converged in $N_C$ that follow $\rho''(0)\sim 1/W_A^2$ (blue solid line is a fit). The  open symbols are not yet converged, [though they are close to converged as depicted in (b)] and we show their $N_C$ dependence, with a fit to the largest $N_C=2^{16}$ (grey line) yielding $\rho''(0)\sim 1/W_A^{2.5}$.
 (Inset) The location of the leading AQCPs $W_A(W_Q)$ are shown as black symbols while the location of the subdominant peak is shown as magenta circles.
 (b) Dependence of the peak in $\rho''(0)$ on the KPM expansion order $N_C$ as we approach the MAT along the line of AQCPs. At the MAT ($W_D=0$) we find this diverges like $\rho''(0)\sim (N_C)^{2.5}$ (fit to the largest three expansion orders shown as a black line).
 All data here is obtained for a system size $L=89$.
 }
 \label{fig:criticalrho2}
\end{figure}

Focusing on $\rho''(0)$ along the avoided line we plot it versus $W_A(W_Q)$ as we approach the MAT in Fig.~\ref{fig:criticalrho2}(a) for both the dominant peak and the subleading peak  while the inset shows the location of each  maximum  in $\rho''(0)$. As we previously described, the dominant peak we are able to converge in $N_C$ when we are far enough away from the MAT. This regime, which is controlled by the first miniband, is well described by  the partial power law
\begin{equation}
    \rho''(0)\sim \frac{1}{(W_A)^{2}} \quad \text{in miniband 1,}
\end{equation} 
and this also describes the  well converged peak (that is also associated with miniband one). We pause here, to note that if this power law where to hold all the way to $W_A=0$ we would in fact find our results to not be internally consistent. To understand why consider the following: we established in Eq.~\eqref{eqn:WA} that $W_A \sim v(W_Q)$, but this implies that the density of states diverges like $\rho''(0) \sim 1/v(W_Q)^2$, a slower divergence than in the clean limit $W_D=0$ where $\rho''(0) \sim 1/v(W_Q)^3$, a contradiction. This issue is alleviated, however, by considering how the second miniband enahances $\rho''(0)$.

The nature in which $\rho''(0)$ is increasing in the second miniband on the other hand is stronger, where our limited numerical data yields  the partial power law
\begin{equation}
    \rho''(0)\sim \frac{1}{(W_A)^{2.5}} \quad \text{in miniband 2}.
\end{equation} 
Importantly, our results are now internally consistent with the limit of $W_D=0$.

To look at this divergence in a different way we consider $\rho''(0)$ as a function of $N_C$ along $W_A$ from where we can, to where we can't converge $\rho''(0)$.  Precisely at the MAT $W_D=0, W_Q=0.6345t$ we find $\rho''(0)$ diverges with $N_C$  as
$\rho''(0)\sim (N_C)^{2.5}$ for the largest $N_C$ that we have accessed.

This brings us to argue that as we approach the MAT each miniband produces a   partial power law like divergence of $\rho''(0)$ with $1/(W_A)$ that is described by
\begin{equation}
    \rho''(0)\sim \frac{1}{(W_{A})^{\beta_n}}
\end{equation}
where $\beta_n$ depends on the $n$th miniband. As two of us conjectured in Ref.~\cite{fuMagicangleSemimetals2020}, there should be an infinite sequence of minibands opening up as we get exponentially closer to the MAT that each correspond to a given order in perturbation theory that is dictated by the irrational nature of the incommensurate wavevector $Q$ in Eq.~\eqref{eqn:QP}. Here it is the sequence $F_{3n}/2$, where $F_{m}$ are Fibonacci numbers; the sequence represents the denominators of the continued fraction of $\sqrt{5}$. As a result, we conjecture that along the line of AQCPs there is an infinite sequence of $\beta_n$'s, obtaining $\beta_3$ in our problem though remains a challenging computational task.

\section{Discussion}
\label{sec:discussion}

In this manuscript, we made a direct link between disorder-driven avoided quantum criticality and the semimetal-to-diffusive metal magic-angle phase transitions tuned by quasiperiodicity. 
By viewing the problem as adding disorder to the quasiperiodic model, we constructed a complete phase diagram. 
The quasiperiodic potential renormalizes the Weyl semimetal parameters and away from the magic-angle transitions the Weyl semimetal survives. 
Adding disorder to this system fills in the band gaps and pseudogaps due to quasiperiodicity, introduces a finite density of states at the Weyl node due to rare regions of the random potential, and rounds out the magic-angle transition into a crossover. 
The line of crossovers is parameterized by the Weyl cone velocity renormalized by the quasiperiodic potential. 
Last, the divergence of the Weyl velocity at the magic-angle transition was computed accurately by approaching the transition along the lines of avoided critical points.

The disordered and quasiperiodic Weyl semimetal model we expect can be realized in future realizations of ultracold atom experiments that use 3D spin-orbit coupling to realize a Weyl semimetal phase. 
Disorder can be introduced using several approaches (e.g., speckle patterns~\cite{goodman2007speckle}, programmable potentials~\cite{RevModPhys.93.025001,PhysRevLett.126.040603}, digital mirror devices~\cite{stuart2018single}), 
while quasiperiodicity can be achieved through a second optical lattice incommensurate with the first. 
The phase transition and cross-overs can be measured through time of flight imaging of wave packet dynamics~\cite{lett_observation_1988} as well as through the spectral function measured using radiofrequency spectroscopy~\cite{gupta_radio-frequency_2003,PhysRevB.106.195123}. 
It will be exciting to see if the transition and its connection to avoided quantum criticality can be exposed in future experiments. Alternatively, circuit quantum electrodynamic setups have been proposed by one of us to also be able to realize this phenomena and see its effect in spectroscopic transport measurements of the junctions~\cite{herrig2022quasiperiodic, herrig2023emulating}.

Future theoretical work needs to utilize a graphical processing units (GPUs) implementation of the KPM to make further progress. To see further minibands open up as we get exponentially close to the magic-angle transition requires systematically increasing the system size beyond $L=89$ as we have considered here that reveals the second miniband at these expansion orders.
In fact, the sequence of perturbation theory order needed to see the next gap opening~\cite{fuMagicangleSemimetals2020} up gives us a clue for this: If we take $L = 3,5$, the first miniband has been opened up, but the unit cell size is the entire size of the system (hence cannot be disordered); for $L =13, 21$, the second miniband has formed and the first miniband can begin to be disordered. For $L = 55,89$ the second miniband can be disordered, while the third miniband has only formed as one unit cell. (Note: We skip even numbers since this degenerate case puts the new miniband gap precisely at zero energy, and no new miniband has yet been formed.) Following this pattern, in order to begin to see disorder effects in the third miniband, we would need to access $L=233, 377$ (that are accessible with GPU implementations of the KPM). 
While continuing this process indefinitely will quickly lead to prohibitively large system sizes, this may also be improved by a renormalization scheme when $W_D/t \ll 1$ that focuses computational effort on the lowest available miniband, which  we leave for future work.

\acknowledgements{We thank Sankar Das Sarma, Sarang Gopalakrishnan, and Elio K\"onig for useful discussions. J.H.P.\ is partially supported  by NSF CAREER Grant No. DMR-1941569 and the Alfred P. Sloan Foundation through a Sloan Research Fellowship.  Part of this work was performed at the Aspen Center for Physics, which is supported by National Science Foundation grant PHY-2210452 (J.H.W, J.H.P.) as well as the Kavli Institute of Theoretical Physics that is supported in part by the National Science Foundation under Grants No. NSF PHY-1748958 and PHY-2309135 (J.H.W, J.H.P.). D.A.H. was supported in part by NSF QLCI grant OMA-2120757. 
The authors acknowledge the following research computing resources: the Beowulf cluster at the Department of Physics and Astronomy of Rutgers University, and the Amarel cluster from the Office of Advanced Research Computing (OARC) at Rutgers, The State University of New Jersey (https://it.rutgers.edu/oarc).  }

\bibliography{references_zotero-2}

\begin{thebibliography}{66}%
\makeatletter
\providecommand \@ifxundefined [1]{%
 \@ifx{#1\undefined}
}%
\providecommand \@ifnum [1]{%
 \ifnum #1\expandafter \@firstoftwo
 \else \expandafter \@secondoftwo
 \fi
}%
\providecommand \@ifx [1]{%
 \ifx #1\expandafter \@firstoftwo
 \else \expandafter \@secondoftwo
 \fi
}%
\providecommand \natexlab [1]{#1}%
\providecommand \enquote  [1]{``#1''}%
\providecommand \bibnamefont  [1]{#1}%
\providecommand \bibfnamefont [1]{#1}%
\providecommand \citenamefont [1]{#1}%
\providecommand \href@noop [0]{\@secondoftwo}%
\providecommand \href [0]{\begingroup \@sanitize@url \@href}%
\providecommand \@href[1]{\@@startlink{#1}\@@href}%
\providecommand \@@href[1]{\endgroup#1\@@endlink}%
\providecommand \@sanitize@url [0]{\catcode `\\12\catcode `\$12\catcode
  `\&12\catcode `\#12\catcode `\^12\catcode `\_12\catcode `\%12\relax}%
\providecommand \@@startlink[1]{}%
\providecommand \@@endlink[0]{}%
\providecommand \url  [0]{\begingroup\@sanitize@url \@url }%
\providecommand \@url [1]{\endgroup\@href {#1}{\urlprefix }}%
\providecommand \urlprefix  [0]{URL }%
\providecommand \Eprint [0]{\href }%
\providecommand \doibase [0]{https://doi.org/}%
\providecommand \selectlanguage [0]{\@gobble}%
\providecommand \bibinfo  [0]{\@secondoftwo}%
\providecommand \bibfield  [0]{\@secondoftwo}%
\providecommand \translation [1]{[#1]}%
\providecommand \BibitemOpen [0]{}%
\providecommand \bibitemStop [0]{}%
\providecommand \bibitemNoStop [0]{.\EOS\space}%
\providecommand \EOS [0]{\spacefactor3000\relax}%
\providecommand \BibitemShut  [1]{\csname bibitem#1\endcsname}%
\let\auto@bib@innerbib\@empty
\bibitem [{\citenamefont {Borisenko}\ \emph {et~al.}(2014)\citenamefont
  {Borisenko}, \citenamefont {Gibson}, \citenamefont {Evtushinsky},
  \citenamefont {Zabolotnyy}, \citenamefont {B{\"u}chner},\ and\ \citenamefont
  {Cava}}]{borisenkoExperimentalRealizationThreedimensional2014}%
  \BibitemOpen
  \bibfield  {author} {\bibinfo {author} {\bibfnamefont {S.}~\bibnamefont
  {Borisenko}}, \bibinfo {author} {\bibfnamefont {Q.}~\bibnamefont {Gibson}},
  \bibinfo {author} {\bibfnamefont {D.}~\bibnamefont {Evtushinsky}}, \bibinfo
  {author} {\bibfnamefont {V.}~\bibnamefont {Zabolotnyy}}, \bibinfo {author}
  {\bibfnamefont {B.}~\bibnamefont {B{\"u}chner}},\ and\ \bibinfo {author}
  {\bibfnamefont {R.~J.}\ \bibnamefont {Cava}},\ }\bibfield  {title} {\bibinfo
  {title} {Experimental realization of a three-dimensional {{Dirac}}
  semimetal},\ }\href {https://doi.org/10.1103/PhysRevLett.113.027603}
  {\bibfield  {journal} {\bibinfo  {journal} {Phys. Rev. Lett.}\ }\textbf
  {\bibinfo {volume} {113}},\ \bibinfo {pages} {027603} (\bibinfo {year}
  {2014})}\BibitemShut {NoStop}%
\bibitem [{\citenamefont {Liu}\ \emph {et~al.}(2014{\natexlab{a}})\citenamefont
  {Liu}, \citenamefont {Zhou}, \citenamefont {Zhang}, \citenamefont {Wang},
  \citenamefont {Weng}, \citenamefont {Prabhakaran}, \citenamefont {Mo},
  \citenamefont {Shen}, \citenamefont {Fang}, \citenamefont {Dai} \emph
  {et~al.}}]{liuDiscoveryThreedimensionalTopological2014}%
  \BibitemOpen
  \bibfield  {author} {\bibinfo {author} {\bibfnamefont {Z.}~\bibnamefont
  {Liu}}, \bibinfo {author} {\bibfnamefont {B.}~\bibnamefont {Zhou}}, \bibinfo
  {author} {\bibfnamefont {Y.}~\bibnamefont {Zhang}}, \bibinfo {author}
  {\bibfnamefont {Z.}~\bibnamefont {Wang}}, \bibinfo {author} {\bibfnamefont
  {H.}~\bibnamefont {Weng}}, \bibinfo {author} {\bibfnamefont {D.}~\bibnamefont
  {Prabhakaran}}, \bibinfo {author} {\bibfnamefont {S.-K.}\ \bibnamefont {Mo}},
  \bibinfo {author} {\bibfnamefont {Z.}~\bibnamefont {Shen}}, \bibinfo {author}
  {\bibfnamefont {Z.}~\bibnamefont {Fang}}, \bibinfo {author} {\bibfnamefont
  {X.}~\bibnamefont {Dai}}, \emph {et~al.},\ }\bibfield  {title} {\bibinfo
  {title} {Discovery of a three-dimensional topological {Dirac} semimetal,
  {Na$_3$Bi}},\ }\href {https://doi.org/10.1126/science.1245085} {\bibfield
  {journal} {\bibinfo  {journal} {Science}\ }\textbf {\bibinfo {volume}
  {343}},\ \bibinfo {pages} {864–867} (\bibinfo {year}
  {2014}{\natexlab{a}})}\BibitemShut {NoStop}%
\bibitem [{\citenamefont {Neupane}\ \emph {et~al.}(2014)\citenamefont
  {Neupane}, \citenamefont {Xu}, \citenamefont {Sankar}, \citenamefont
  {Alidoust}, \citenamefont {Bian}, \citenamefont {Liu}, \citenamefont
  {Belopolski}, \citenamefont {Chang}, \citenamefont {Jeng}, \citenamefont
  {Lin} \emph {et~al.}}]{Neupane-2014}%
  \BibitemOpen
  \bibfield  {author} {\bibinfo {author} {\bibfnamefont {M.}~\bibnamefont
  {Neupane}}, \bibinfo {author} {\bibfnamefont {S.-Y.}\ \bibnamefont {Xu}},
  \bibinfo {author} {\bibfnamefont {R.}~\bibnamefont {Sankar}}, \bibinfo
  {author} {\bibfnamefont {N.}~\bibnamefont {Alidoust}}, \bibinfo {author}
  {\bibfnamefont {G.}~\bibnamefont {Bian}}, \bibinfo {author} {\bibfnamefont
  {C.}~\bibnamefont {Liu}}, \bibinfo {author} {\bibfnamefont {I.}~\bibnamefont
  {Belopolski}}, \bibinfo {author} {\bibfnamefont {T.-R.}\ \bibnamefont
  {Chang}}, \bibinfo {author} {\bibfnamefont {H.-T.}\ \bibnamefont {Jeng}},
  \bibinfo {author} {\bibfnamefont {H.}~\bibnamefont {Lin}}, \emph {et~al.},\
  }\bibfield  {title} {\bibinfo {title} {Observation of a three-dimensional
  topological {{Dirac}} semimetal phase in high-mobility
  {{Cd}}{$_{3}$}{{As}}{$_{2}$}},\ }\href {https://doi.org/10.1038/ncomms4786}
  {\bibfield  {journal} {\bibinfo  {journal} {Nat. Commun.}\ }\textbf {\bibinfo
  {volume} {5}},\ \bibinfo {pages} {3786} (\bibinfo {year} {2014})}\BibitemShut
  {NoStop}%
\bibitem [{\citenamefont {Lv}\ \emph {et~al.}(2015{\natexlab{a}})\citenamefont
  {Lv}, \citenamefont {Weng}, \citenamefont {Fu}, \citenamefont {Wang},
  \citenamefont {Miao}, \citenamefont {Ma}, \citenamefont {Richard},
  \citenamefont {Huang}, \citenamefont {Zhao}, \citenamefont {Chen},
  \citenamefont {Fang}, \citenamefont {Dai}, \citenamefont {Qian},\ and\
  \citenamefont {Ding}}]{lvExperimentalDiscoveryWeyl2015}%
  \BibitemOpen
  \bibfield  {author} {\bibinfo {author} {\bibfnamefont {B.~Q.}\ \bibnamefont
  {Lv}}, \bibinfo {author} {\bibfnamefont {H.~M.}\ \bibnamefont {Weng}},
  \bibinfo {author} {\bibfnamefont {B.~B.}\ \bibnamefont {Fu}}, \bibinfo
  {author} {\bibfnamefont {X.~P.}\ \bibnamefont {Wang}}, \bibinfo {author}
  {\bibfnamefont {H.}~\bibnamefont {Miao}}, \bibinfo {author} {\bibfnamefont
  {J.}~\bibnamefont {Ma}}, \bibinfo {author} {\bibfnamefont {P.}~\bibnamefont
  {Richard}}, \bibinfo {author} {\bibfnamefont {X.~C.}\ \bibnamefont {Huang}},
  \bibinfo {author} {\bibfnamefont {L.~X.}\ \bibnamefont {Zhao}}, \bibinfo
  {author} {\bibfnamefont {G.~F.}\ \bibnamefont {Chen}}, \bibinfo {author}
  {\bibfnamefont {Z.}~\bibnamefont {Fang}}, \bibinfo {author} {\bibfnamefont
  {X.}~\bibnamefont {Dai}}, \bibinfo {author} {\bibfnamefont {T.}~\bibnamefont
  {Qian}},\ and\ \bibinfo {author} {\bibfnamefont {H.}~\bibnamefont {Ding}},\
  }\bibfield  {title} {\bibinfo {title} {Experimental discovery of {{Weyl}}
  semimetal {{TaAs}}},\ }\href {https://doi.org/10.1103/PhysRevX.5.031013}
  {\bibfield  {journal} {\bibinfo  {journal} {Phys. Rev. X}\ }\textbf {\bibinfo
  {volume} {5}},\ \bibinfo {pages} {031013} (\bibinfo {year}
  {2015}{\natexlab{a}})}\BibitemShut {NoStop}%
\bibitem [{\citenamefont {Lv}\ \emph {et~al.}(2015{\natexlab{b}})\citenamefont
  {Lv}, \citenamefont {Xu}, \citenamefont {Weng}, \citenamefont {Ma},
  \citenamefont {Richard}, \citenamefont {Huang}, \citenamefont {Zhao},
  \citenamefont {Chen}, \citenamefont {Matt}, \citenamefont {Bisti},
  \citenamefont {Strocov}, \citenamefont {Mesot}, \citenamefont {Fang},
  \citenamefont {Dai}, \citenamefont {Qian}, \citenamefont {Shi},\ and\
  \citenamefont {Ding}}]{lv_observation_2015}%
  \BibitemOpen
  \bibfield  {author} {\bibinfo {author} {\bibfnamefont {B.~Q.}\ \bibnamefont
  {Lv}}, \bibinfo {author} {\bibfnamefont {N.}~\bibnamefont {Xu}}, \bibinfo
  {author} {\bibfnamefont {H.~M.}\ \bibnamefont {Weng}}, \bibinfo {author}
  {\bibfnamefont {J.~Z.}\ \bibnamefont {Ma}}, \bibinfo {author} {\bibfnamefont
  {P.}~\bibnamefont {Richard}}, \bibinfo {author} {\bibfnamefont {X.~C.}\
  \bibnamefont {Huang}}, \bibinfo {author} {\bibfnamefont {L.~X.}\ \bibnamefont
  {Zhao}}, \bibinfo {author} {\bibfnamefont {G.~F.}\ \bibnamefont {Chen}},
  \bibinfo {author} {\bibfnamefont {C.~E.}\ \bibnamefont {Matt}}, \bibinfo
  {author} {\bibfnamefont {F.}~\bibnamefont {Bisti}}, \bibinfo {author}
  {\bibfnamefont {V.~N.}\ \bibnamefont {Strocov}}, \bibinfo {author}
  {\bibfnamefont {J.}~\bibnamefont {Mesot}}, \bibinfo {author} {\bibfnamefont
  {Z.}~\bibnamefont {Fang}}, \bibinfo {author} {\bibfnamefont {X.}~\bibnamefont
  {Dai}}, \bibinfo {author} {\bibfnamefont {T.}~\bibnamefont {Qian}}, \bibinfo
  {author} {\bibfnamefont {M.}~\bibnamefont {Shi}},\ and\ \bibinfo {author}
  {\bibfnamefont {H.}~\bibnamefont {Ding}},\ }\bibfield  {title} {\bibinfo
  {title} {Observation of {{Weyl}} nodes in {{TaAs}}},\ }\href
  {https://doi.org/10.1038/nphys3426} {\bibfield  {journal} {\bibinfo
  {journal} {Nature Phys}\ }\textbf {\bibinfo {volume} {11}},\ \bibinfo {pages}
  {724} (\bibinfo {year} {2015}{\natexlab{b}})}\BibitemShut {NoStop}%
\bibitem [{\citenamefont {Xu}\ \emph {et~al.}(2015{\natexlab{a}})\citenamefont
  {Xu}, \citenamefont {Belopolski}, \citenamefont {Alidoust}, \citenamefont
  {Neupane}, \citenamefont {Bian}, \citenamefont {Zhang}, \citenamefont
  {Sankar}, \citenamefont {Chang}, \citenamefont {Yuan}, \citenamefont {Lee}
  \emph {et~al.}}]{Xu3-2015}%
  \BibitemOpen
  \bibfield  {author} {\bibinfo {author} {\bibfnamefont {S.-Y.}\ \bibnamefont
  {Xu}}, \bibinfo {author} {\bibfnamefont {I.}~\bibnamefont {Belopolski}},
  \bibinfo {author} {\bibfnamefont {N.}~\bibnamefont {Alidoust}}, \bibinfo
  {author} {\bibfnamefont {M.}~\bibnamefont {Neupane}}, \bibinfo {author}
  {\bibfnamefont {G.}~\bibnamefont {Bian}}, \bibinfo {author} {\bibfnamefont
  {C.}~\bibnamefont {Zhang}}, \bibinfo {author} {\bibfnamefont
  {R.}~\bibnamefont {Sankar}}, \bibinfo {author} {\bibfnamefont
  {G.}~\bibnamefont {Chang}}, \bibinfo {author} {\bibfnamefont
  {Z.}~\bibnamefont {Yuan}}, \bibinfo {author} {\bibfnamefont {C.-C.}\
  \bibnamefont {Lee}}, \emph {et~al.},\ }\bibfield  {title} {\bibinfo {title}
  {Discovery of a {{Weyl}} fermion semimetal and topological {{Fermi}} arcs},\
  }\href {https://doi.org/10.1126/science.aaa9297} {\bibfield  {journal}
  {\bibinfo  {journal} {Science}\ }\textbf {\bibinfo {volume} {349}},\ \bibinfo
  {pages} {613} (\bibinfo {year} {2015}{\natexlab{a}})}\BibitemShut {NoStop}%
\bibitem [{\citenamefont {Xu}\ \emph {et~al.}(2015{\natexlab{b}})\citenamefont
  {Xu}, \citenamefont {Alidoust}, \citenamefont {Belopolski}, \citenamefont
  {Yuan}, \citenamefont {Bian}, \citenamefont {Chang}, \citenamefont {Zheng},
  \citenamefont {Strocov}, \citenamefont {Sanchez}, \citenamefont {Chang} \emph
  {et~al.}}]{Xu2-2015}%
  \BibitemOpen
  \bibfield  {author} {\bibinfo {author} {\bibfnamefont {S.-Y.}\ \bibnamefont
  {Xu}}, \bibinfo {author} {\bibfnamefont {N.}~\bibnamefont {Alidoust}},
  \bibinfo {author} {\bibfnamefont {I.}~\bibnamefont {Belopolski}}, \bibinfo
  {author} {\bibfnamefont {Z.}~\bibnamefont {Yuan}}, \bibinfo {author}
  {\bibfnamefont {G.}~\bibnamefont {Bian}}, \bibinfo {author} {\bibfnamefont
  {T.-R.}\ \bibnamefont {Chang}}, \bibinfo {author} {\bibfnamefont
  {H.}~\bibnamefont {Zheng}}, \bibinfo {author} {\bibfnamefont {V.~N.}\
  \bibnamefont {Strocov}}, \bibinfo {author} {\bibfnamefont {D.~S.}\
  \bibnamefont {Sanchez}}, \bibinfo {author} {\bibfnamefont {G.}~\bibnamefont
  {Chang}}, \emph {et~al.},\ }\bibfield  {title} {\bibinfo {title} {Discovery
  of a {{Weyl}} fermion state with {{Fermi}} arcs in niobium arsenide},\ }\href
  {https://doi.org/10.1038/nphys3437} {\bibfield  {journal} {\bibinfo
  {journal} {Nat. Phys.}\ ,\ \bibinfo {pages} {748}} (\bibinfo {year}
  {2015}{\natexlab{b}})}\BibitemShut {NoStop}%
\bibitem [{\citenamefont {Xu}\ \emph {et~al.}(2015{\natexlab{c}})\citenamefont
  {Xu}, \citenamefont {Belopolski}, \citenamefont {Sanchez}, \citenamefont
  {Zhang}, \citenamefont {Chang}, \citenamefont {Guo}, \citenamefont {Bian},
  \citenamefont {Yuan}, \citenamefont {Lu}, \citenamefont {Chang},
  \citenamefont {Shibayev}, \citenamefont {Prokopovych}, \citenamefont
  {Alidoust}, \citenamefont {Zheng}, \citenamefont {Lee}, \citenamefont
  {Huang}, \citenamefont {Sankar}, \citenamefont {Chou}, \citenamefont {Hsu},
  \citenamefont {Jeng}, \citenamefont {Bansil}, \citenamefont {Neupert},
  \citenamefont {Strocov}, \citenamefont {Lin}, \citenamefont {Jia},\ and\
  \citenamefont {Hasan}}]{xuExperimentalDiscoveryTopological2015}%
  \BibitemOpen
  \bibfield  {author} {\bibinfo {author} {\bibfnamefont {S.-Y.}\ \bibnamefont
  {Xu}}, \bibinfo {author} {\bibfnamefont {I.}~\bibnamefont {Belopolski}},
  \bibinfo {author} {\bibfnamefont {D.~S.}\ \bibnamefont {Sanchez}}, \bibinfo
  {author} {\bibfnamefont {C.}~\bibnamefont {Zhang}}, \bibinfo {author}
  {\bibfnamefont {G.}~\bibnamefont {Chang}}, \bibinfo {author} {\bibfnamefont
  {C.}~\bibnamefont {Guo}}, \bibinfo {author} {\bibfnamefont {G.}~\bibnamefont
  {Bian}}, \bibinfo {author} {\bibfnamefont {Z.}~\bibnamefont {Yuan}}, \bibinfo
  {author} {\bibfnamefont {H.}~\bibnamefont {Lu}}, \bibinfo {author}
  {\bibfnamefont {T.-R.}\ \bibnamefont {Chang}}, \bibinfo {author}
  {\bibfnamefont {P.~P.}\ \bibnamefont {Shibayev}}, \bibinfo {author}
  {\bibfnamefont {M.~L.}\ \bibnamefont {Prokopovych}}, \bibinfo {author}
  {\bibfnamefont {N.}~\bibnamefont {Alidoust}}, \bibinfo {author}
  {\bibfnamefont {H.}~\bibnamefont {Zheng}}, \bibinfo {author} {\bibfnamefont
  {C.-C.}\ \bibnamefont {Lee}}, \bibinfo {author} {\bibfnamefont {S.-M.}\
  \bibnamefont {Huang}}, \bibinfo {author} {\bibfnamefont {R.}~\bibnamefont
  {Sankar}}, \bibinfo {author} {\bibfnamefont {F.}~\bibnamefont {Chou}},
  \bibinfo {author} {\bibfnamefont {C.-H.}\ \bibnamefont {Hsu}}, \bibinfo
  {author} {\bibfnamefont {H.-T.}\ \bibnamefont {Jeng}}, \bibinfo {author}
  {\bibfnamefont {A.}~\bibnamefont {Bansil}}, \bibinfo {author} {\bibfnamefont
  {T.}~\bibnamefont {Neupert}}, \bibinfo {author} {\bibfnamefont {V.~N.}\
  \bibnamefont {Strocov}}, \bibinfo {author} {\bibfnamefont {H.}~\bibnamefont
  {Lin}}, \bibinfo {author} {\bibfnamefont {S.}~\bibnamefont {Jia}},\ and\
  \bibinfo {author} {\bibfnamefont {M.~Z.}\ \bibnamefont {Hasan}},\ }\bibfield
  {title} {\bibinfo {title} {Experimental discovery of a topological {{Weyl}}
  semimetal state in {{TaP}}},\ }\href {https://doi.org/10.1126/sciadv.1501092}
  {\bibfield  {journal} {\bibinfo  {journal} {Sci. Adv.}\ }\textbf {\bibinfo
  {volume} {1}},\ \bibinfo {pages} {e1501092} (\bibinfo {year}
  {2015}{\natexlab{c}})}\BibitemShut {NoStop}%
\bibitem [{\citenamefont {Xu}\ \emph {et~al.}(2016)\citenamefont {Xu},
  \citenamefont {Weng}, \citenamefont {Lv}, \citenamefont {Matt}, \citenamefont
  {Park}, \citenamefont {Bisti}, \citenamefont {Strocov}, \citenamefont
  {Gawryluk}, \citenamefont {Pomjakushina}, \citenamefont {Conder},
  \citenamefont {Plumb}, \citenamefont {Radovic}, \citenamefont {Aut{\`e}s},
  \citenamefont {Yazyev}, \citenamefont {Fang}, \citenamefont {Dai},
  \citenamefont {Qian}, \citenamefont {Mesot}, \citenamefont {Ding},\ and\
  \citenamefont {Shi}}]{xuObservationWeylNodes2016}%
  \BibitemOpen
  \bibfield  {author} {\bibinfo {author} {\bibfnamefont {N.}~\bibnamefont
  {Xu}}, \bibinfo {author} {\bibfnamefont {H.~M.}\ \bibnamefont {Weng}},
  \bibinfo {author} {\bibfnamefont {B.~Q.}\ \bibnamefont {Lv}}, \bibinfo
  {author} {\bibfnamefont {C.~E.}\ \bibnamefont {Matt}}, \bibinfo {author}
  {\bibfnamefont {J.}~\bibnamefont {Park}}, \bibinfo {author} {\bibfnamefont
  {F.}~\bibnamefont {Bisti}}, \bibinfo {author} {\bibfnamefont {V.~N.}\
  \bibnamefont {Strocov}}, \bibinfo {author} {\bibfnamefont {D.}~\bibnamefont
  {Gawryluk}}, \bibinfo {author} {\bibfnamefont {E.}~\bibnamefont
  {Pomjakushina}}, \bibinfo {author} {\bibfnamefont {K.}~\bibnamefont
  {Conder}}, \bibinfo {author} {\bibfnamefont {N.~C.}\ \bibnamefont {Plumb}},
  \bibinfo {author} {\bibfnamefont {M.}~\bibnamefont {Radovic}}, \bibinfo
  {author} {\bibfnamefont {G.}~\bibnamefont {Aut{\`e}s}}, \bibinfo {author}
  {\bibfnamefont {O.~V.}\ \bibnamefont {Yazyev}}, \bibinfo {author}
  {\bibfnamefont {Z.}~\bibnamefont {Fang}}, \bibinfo {author} {\bibfnamefont
  {X.}~\bibnamefont {Dai}}, \bibinfo {author} {\bibfnamefont {T.}~\bibnamefont
  {Qian}}, \bibinfo {author} {\bibfnamefont {J.}~\bibnamefont {Mesot}},
  \bibinfo {author} {\bibfnamefont {H.}~\bibnamefont {Ding}},\ and\ \bibinfo
  {author} {\bibfnamefont {M.}~\bibnamefont {Shi}},\ }\bibfield  {title}
  {\bibinfo {title} {Observation of {{Weyl}} nodes and {{Fermi}} arcs in
  tantalum phosphide},\ }\href {https://doi.org/10.1038/ncomms11006} {\bibfield
   {journal} {\bibinfo  {journal} {Nat. Commun.}\ }\textbf {\bibinfo {volume}
  {7}},\ \bibinfo {pages} {11006} (\bibinfo {year} {2016})}\BibitemShut
  {NoStop}%
\bibitem [{\citenamefont {Tafti}\ \emph {et~al.}(2012)\citenamefont {Tafti},
  \citenamefont {Ishikawa}, \citenamefont {McCollam}, \citenamefont
  {Nakatsuji},\ and\ \citenamefont {Julian}}]{tafti_pressure-tuned_2012}%
  \BibitemOpen
  \bibfield  {author} {\bibinfo {author} {\bibfnamefont {F.~F.}\ \bibnamefont
  {Tafti}}, \bibinfo {author} {\bibfnamefont {J.~J.}\ \bibnamefont {Ishikawa}},
  \bibinfo {author} {\bibfnamefont {A.}~\bibnamefont {McCollam}}, \bibinfo
  {author} {\bibfnamefont {S.}~\bibnamefont {Nakatsuji}},\ and\ \bibinfo
  {author} {\bibfnamefont {S.~R.}\ \bibnamefont {Julian}},\ }\bibfield  {title}
  {\bibinfo {title} {Pressure-tuned insulator to metal transition in {{Eu}} 2
  {{Ir}} 2 {{O}} 7},\ }\href {https://doi.org/10.1103/PhysRevB.85.205104}
  {\bibfield  {journal} {\bibinfo  {journal} {Phys. Rev. B}\ }\textbf {\bibinfo
  {volume} {85}},\ \bibinfo {pages} {205104} (\bibinfo {year}
  {2012})}\BibitemShut {NoStop}%
\bibitem [{\citenamefont {Sushkov}\ \emph {et~al.}(2015)\citenamefont
  {Sushkov}, \citenamefont {Hofmann}, \citenamefont {Jenkins}, \citenamefont
  {Ishikawa}, \citenamefont {Nakatsuji}, \citenamefont {Das~Sarma},\ and\
  \citenamefont {Drew}}]{sushkov_optical_2015}%
  \BibitemOpen
  \bibfield  {author} {\bibinfo {author} {\bibfnamefont {A.~B.}\ \bibnamefont
  {Sushkov}}, \bibinfo {author} {\bibfnamefont {J.~B.}\ \bibnamefont
  {Hofmann}}, \bibinfo {author} {\bibfnamefont {G.~S.}\ \bibnamefont
  {Jenkins}}, \bibinfo {author} {\bibfnamefont {J.}~\bibnamefont {Ishikawa}},
  \bibinfo {author} {\bibfnamefont {S.}~\bibnamefont {Nakatsuji}}, \bibinfo
  {author} {\bibfnamefont {S.}~\bibnamefont {Das~Sarma}},\ and\ \bibinfo
  {author} {\bibfnamefont {H.~D.}\ \bibnamefont {Drew}},\ }\bibfield  {title}
  {\bibinfo {title} {Optical evidence for a {{Weyl}} semimetal state in
  pyrochlore {{Eu}} 2 {{Ir}} 2 {{O}} 7},\ }\href
  {https://doi.org/10.1103/PhysRevB.92.241108} {\bibfield  {journal} {\bibinfo
  {journal} {Phys. Rev. B}\ }\textbf {\bibinfo {volume} {92}},\ \bibinfo
  {pages} {241108} (\bibinfo {year} {2015})}\BibitemShut {NoStop}%
\bibitem [{\citenamefont {Lai}\ \emph {et~al.}(2018)\citenamefont {Lai},
  \citenamefont {Grefe}, \citenamefont {Paschen},\ and\ \citenamefont
  {Si}}]{laiWeylKondoSemimetal2018}%
  \BibitemOpen
  \bibfield  {author} {\bibinfo {author} {\bibfnamefont {H.-H.}\ \bibnamefont
  {Lai}}, \bibinfo {author} {\bibfnamefont {S.~E.}\ \bibnamefont {Grefe}},
  \bibinfo {author} {\bibfnamefont {S.}~\bibnamefont {Paschen}},\ and\ \bibinfo
  {author} {\bibfnamefont {Q.}~\bibnamefont {Si}},\ }\bibfield  {title}
  {\bibinfo {title} {Weyl\textendash{{Kondo}} semimetal in heavy-fermion
  systems},\ }\href {https://doi.org/10.1073/pnas.1715851115} {\bibfield
  {journal} {\bibinfo  {journal} {Proc. Natl. Acad. Sci. U.S.A.}\ }\textbf
  {\bibinfo {volume} {115}},\ \bibinfo {pages} {93} (\bibinfo {year}
  {2018})}\BibitemShut {NoStop}%
\bibitem [{\citenamefont {Chang}\ and\ \citenamefont
  {Coleman}(2018)}]{changParityviolatingHybridizationHeavy2018}%
  \BibitemOpen
  \bibfield  {author} {\bibinfo {author} {\bibfnamefont {P.-Y.}\ \bibnamefont
  {Chang}}\ and\ \bibinfo {author} {\bibfnamefont {P.}~\bibnamefont
  {Coleman}},\ }\bibfield  {title} {\bibinfo {title} {Parity-violating
  hybridization in heavy {{Weyl}} semimetals},\ }\href
  {https://doi.org/10.1103/PhysRevB.97.155134} {\bibfield  {journal} {\bibinfo
  {journal} {Phys. Rev. B}\ }\textbf {\bibinfo {volume} {97}},\ \bibinfo
  {pages} {155134} (\bibinfo {year} {2018})}\BibitemShut {NoStop}%
\bibitem [{\citenamefont {Telang}\ \emph {et~al.}(2019)\citenamefont {Telang},
  \citenamefont {Mishra}, \citenamefont {Prando}, \citenamefont {Sood},\ and\
  \citenamefont {Singh}}]{telang_anomalous_2019}%
  \BibitemOpen
  \bibfield  {author} {\bibinfo {author} {\bibfnamefont {P.}~\bibnamefont
  {Telang}}, \bibinfo {author} {\bibfnamefont {K.}~\bibnamefont {Mishra}},
  \bibinfo {author} {\bibfnamefont {G.}~\bibnamefont {Prando}}, \bibinfo
  {author} {\bibfnamefont {A.~K.}\ \bibnamefont {Sood}},\ and\ \bibinfo
  {author} {\bibfnamefont {S.}~\bibnamefont {Singh}},\ }\bibfield  {title}
  {\bibinfo {title} {Anomalous lattice contraction and emergent electronic
  phases in {{Bi-doped Eu}} 2 {{Ir}} 2 {{O}} 7},\ }\href
  {https://doi.org/10.1103/PhysRevB.99.201112} {\bibfield  {journal} {\bibinfo
  {journal} {Phys. Rev. B}\ }\textbf {\bibinfo {volume} {99}},\ \bibinfo
  {pages} {201112} (\bibinfo {year} {2019})}\BibitemShut {NoStop}%
\bibitem [{\citenamefont {Dzsaber}\ \emph {et~al.}(2019)\citenamefont
  {Dzsaber}, \citenamefont {Yan}, \citenamefont {Taupin}, \citenamefont
  {Eguchi}, \citenamefont {Prokofiev}, \citenamefont {Shiroka}, \citenamefont
  {Blaha}, \citenamefont {Rubel}, \citenamefont {Grefe}, \citenamefont {Lai},
  \citenamefont {Si},\ and\ \citenamefont
  {Paschen}}]{dzsaberGiantSpontaneousHall2019}%
  \BibitemOpen
  \bibfield  {author} {\bibinfo {author} {\bibfnamefont {S.}~\bibnamefont
  {Dzsaber}}, \bibinfo {author} {\bibfnamefont {X.}~\bibnamefont {Yan}},
  \bibinfo {author} {\bibfnamefont {M.}~\bibnamefont {Taupin}}, \bibinfo
  {author} {\bibfnamefont {G.}~\bibnamefont {Eguchi}}, \bibinfo {author}
  {\bibfnamefont {A.}~\bibnamefont {Prokofiev}}, \bibinfo {author}
  {\bibfnamefont {T.}~\bibnamefont {Shiroka}}, \bibinfo {author} {\bibfnamefont
  {P.}~\bibnamefont {Blaha}}, \bibinfo {author} {\bibfnamefont
  {O.}~\bibnamefont {Rubel}}, \bibinfo {author} {\bibfnamefont {S.~E.}\
  \bibnamefont {Grefe}}, \bibinfo {author} {\bibfnamefont {H.-H.}\ \bibnamefont
  {Lai}}, \bibinfo {author} {\bibfnamefont {Q.}~\bibnamefont {Si}},\ and\
  \bibinfo {author} {\bibfnamefont {S.}~\bibnamefont {Paschen}},\ }\bibfield
  {title} {\bibinfo {title} {Giant spontaneous {{Hall}} effect in a nonmagnetic
  {{Weyl-Kondo}} semimetal},\ }\href@noop {} {\bibfield  {journal} {\bibinfo
  {journal} {arXiv:1811.02819 [cond-mat]}\ } (\bibinfo {year} {2019})},\
  \Eprint {https://arxiv.org/abs/1811.02819} {arxiv:1811.02819 [cond-mat]}
  \BibitemShut {NoStop}%
\bibitem [{\citenamefont {Kuroda}\ \emph {et~al.}(2017)\citenamefont {Kuroda},
  \citenamefont {Tomita}, \citenamefont {Suzuki}, \citenamefont {Bareille},
  \citenamefont {Nugroho}, \citenamefont {Goswami}, \citenamefont {Ochi},
  \citenamefont {Ikhlas}, \citenamefont {Nakayama}, \citenamefont {Akebi},
  \citenamefont {Noguchi}, \citenamefont {Ishii}, \citenamefont {Inami},
  \citenamefont {Ono}, \citenamefont {Kumigashira}, \citenamefont {Varykhalov},
  \citenamefont {Muro}, \citenamefont {Koretsune}, \citenamefont {Arita},
  \citenamefont {Shin}, \citenamefont {Kondo},\ and\ \citenamefont
  {Nakatsuji}}]{kurodaEvidenceMagneticWeyl2017}%
  \BibitemOpen
  \bibfield  {author} {\bibinfo {author} {\bibfnamefont {K.}~\bibnamefont
  {Kuroda}}, \bibinfo {author} {\bibfnamefont {T.}~\bibnamefont {Tomita}},
  \bibinfo {author} {\bibfnamefont {M.-T.}\ \bibnamefont {Suzuki}}, \bibinfo
  {author} {\bibfnamefont {C.}~\bibnamefont {Bareille}}, \bibinfo {author}
  {\bibfnamefont {A.~A.}\ \bibnamefont {Nugroho}}, \bibinfo {author}
  {\bibfnamefont {P.}~\bibnamefont {Goswami}}, \bibinfo {author} {\bibfnamefont
  {M.}~\bibnamefont {Ochi}}, \bibinfo {author} {\bibfnamefont {M.}~\bibnamefont
  {Ikhlas}}, \bibinfo {author} {\bibfnamefont {M.}~\bibnamefont {Nakayama}},
  \bibinfo {author} {\bibfnamefont {S.}~\bibnamefont {Akebi}}, \bibinfo
  {author} {\bibfnamefont {R.}~\bibnamefont {Noguchi}}, \bibinfo {author}
  {\bibfnamefont {R.}~\bibnamefont {Ishii}}, \bibinfo {author} {\bibfnamefont
  {N.}~\bibnamefont {Inami}}, \bibinfo {author} {\bibfnamefont
  {K.}~\bibnamefont {Ono}}, \bibinfo {author} {\bibfnamefont {H.}~\bibnamefont
  {Kumigashira}}, \bibinfo {author} {\bibfnamefont {A.}~\bibnamefont
  {Varykhalov}}, \bibinfo {author} {\bibfnamefont {T.}~\bibnamefont {Muro}},
  \bibinfo {author} {\bibfnamefont {T.}~\bibnamefont {Koretsune}}, \bibinfo
  {author} {\bibfnamefont {R.}~\bibnamefont {Arita}}, \bibinfo {author}
  {\bibfnamefont {S.}~\bibnamefont {Shin}}, \bibinfo {author} {\bibfnamefont
  {T.}~\bibnamefont {Kondo}},\ and\ \bibinfo {author} {\bibfnamefont
  {S.}~\bibnamefont {Nakatsuji}},\ }\bibfield  {title} {\bibinfo {title}
  {Evidence for magnetic {{Weyl}} fermions in a correlated metal},\ }\href
  {https://doi.org/10.1038/nmat4987} {\bibfield  {journal} {\bibinfo  {journal}
  {Nat. Mater.}\ }\textbf {\bibinfo {volume} {16}},\ \bibinfo {pages} {1090}
  (\bibinfo {year} {2017})}\BibitemShut {NoStop}%
\bibitem [{\citenamefont {Guo}\ \emph {et~al.}(2018)\citenamefont {Guo},
  \citenamefont {Wu}, \citenamefont {Wu}, \citenamefont {Smidman},
  \citenamefont {Cao}, \citenamefont {Bostwick}, \citenamefont {Jozwiak},
  \citenamefont {Rotenberg}, \citenamefont {Liu}, \citenamefont {Steglich},\
  and\ \citenamefont {Yuan}}]{guoEvidenceWeylFermions2018}%
  \BibitemOpen
  \bibfield  {author} {\bibinfo {author} {\bibfnamefont {C.~Y.}\ \bibnamefont
  {Guo}}, \bibinfo {author} {\bibfnamefont {F.}~\bibnamefont {Wu}}, \bibinfo
  {author} {\bibfnamefont {Z.~Z.}\ \bibnamefont {Wu}}, \bibinfo {author}
  {\bibfnamefont {M.}~\bibnamefont {Smidman}}, \bibinfo {author} {\bibfnamefont
  {C.}~\bibnamefont {Cao}}, \bibinfo {author} {\bibfnamefont {A.}~\bibnamefont
  {Bostwick}}, \bibinfo {author} {\bibfnamefont {C.}~\bibnamefont {Jozwiak}},
  \bibinfo {author} {\bibfnamefont {E.}~\bibnamefont {Rotenberg}}, \bibinfo
  {author} {\bibfnamefont {Y.}~\bibnamefont {Liu}}, \bibinfo {author}
  {\bibfnamefont {F.}~\bibnamefont {Steglich}},\ and\ \bibinfo {author}
  {\bibfnamefont {H.~Q.}\ \bibnamefont {Yuan}},\ }\bibfield  {title} {\bibinfo
  {title} {Evidence for {{Weyl}} fermions in a canonical heavy-fermion
  semimetal {{YbPtBi}}},\ }\href {https://doi.org/10.1038/s41467-018-06782-1}
  {\bibfield  {journal} {\bibinfo  {journal} {Nat. Commun.}\ }\textbf {\bibinfo
  {volume} {9}},\ \bibinfo {pages} {4622} (\bibinfo {year} {2018})}\BibitemShut
  {NoStop}%
\bibitem [{\citenamefont {Liu}\ \emph {et~al.}(2021)\citenamefont {Liu},
  \citenamefont {Fang}, \citenamefont {Fu}, \citenamefont {Ge}, \citenamefont
  {Kareev}, \citenamefont {Kim}, \citenamefont {Choi}, \citenamefont
  {Karapetrova}, \citenamefont {Zhang}, \citenamefont {Gu}, \citenamefont
  {Choi}, \citenamefont {Wen}, \citenamefont {Wilson}, \citenamefont {Fabbris},
  \citenamefont {Ryan}, \citenamefont {Freeland}, \citenamefont {Haskel},
  \citenamefont {Wu}, \citenamefont {Pixley},\ and\ \citenamefont
  {Chakhalian}}]{liu_magnetic_2021}%
  \BibitemOpen
  \bibfield  {author} {\bibinfo {author} {\bibfnamefont {X.}~\bibnamefont
  {Liu}}, \bibinfo {author} {\bibfnamefont {S.}~\bibnamefont {Fang}}, \bibinfo
  {author} {\bibfnamefont {Y.}~\bibnamefont {Fu}}, \bibinfo {author}
  {\bibfnamefont {W.}~\bibnamefont {Ge}}, \bibinfo {author} {\bibfnamefont
  {M.}~\bibnamefont {Kareev}}, \bibinfo {author} {\bibfnamefont {J.-W.}\
  \bibnamefont {Kim}}, \bibinfo {author} {\bibfnamefont {Y.}~\bibnamefont
  {Choi}}, \bibinfo {author} {\bibfnamefont {E.}~\bibnamefont {Karapetrova}},
  \bibinfo {author} {\bibfnamefont {Q.}~\bibnamefont {Zhang}}, \bibinfo
  {author} {\bibfnamefont {L.}~\bibnamefont {Gu}}, \bibinfo {author}
  {\bibfnamefont {E.-S.}\ \bibnamefont {Choi}}, \bibinfo {author}
  {\bibfnamefont {F.}~\bibnamefont {Wen}}, \bibinfo {author} {\bibfnamefont
  {J.~H.}\ \bibnamefont {Wilson}}, \bibinfo {author} {\bibfnamefont
  {G.}~\bibnamefont {Fabbris}}, \bibinfo {author} {\bibfnamefont {P.~J.}\
  \bibnamefont {Ryan}}, \bibinfo {author} {\bibfnamefont {J.~W.}\ \bibnamefont
  {Freeland}}, \bibinfo {author} {\bibfnamefont {D.}~\bibnamefont {Haskel}},
  \bibinfo {author} {\bibfnamefont {W.}~\bibnamefont {Wu}}, \bibinfo {author}
  {\bibfnamefont {J.~H.}\ \bibnamefont {Pixley}},\ and\ \bibinfo {author}
  {\bibfnamefont {J.}~\bibnamefont {Chakhalian}},\ }\bibfield  {title}
  {\bibinfo {title} {Magnetic {{Weyl Semimetallic Phase}} in {{Thin Films}} of
  {{Eu2Ir2O7}}},\ }\href {https://doi.org/10.1103/PhysRevLett.127.277204}
  {\bibfield  {journal} {\bibinfo  {journal} {Phys. Rev. Lett.}\ }\textbf
  {\bibinfo {volume} {127}},\ \bibinfo {pages} {277204} (\bibinfo {year}
  {2021})},\ \Eprint {https://arxiv.org/abs/2106.04062} {arxiv:2106.04062}
  \BibitemShut {NoStop}%
\bibitem [{\citenamefont {Liu}\ \emph {et~al.}(2014{\natexlab{b}})\citenamefont
  {Liu}, \citenamefont {Jiang}, \citenamefont {Zhou}, \citenamefont {Wang},
  \citenamefont {Zhang}, \citenamefont {Weng}, \citenamefont {Prabhakaran},
  \citenamefont {Mo}, \citenamefont {Peng}, \citenamefont {Dudin} \emph
  {et~al.}}]{liuStableThreedimensionalTopological2014}%
  \BibitemOpen
  \bibfield  {author} {\bibinfo {author} {\bibfnamefont {Z.}~\bibnamefont
  {Liu}}, \bibinfo {author} {\bibfnamefont {J.}~\bibnamefont {Jiang}}, \bibinfo
  {author} {\bibfnamefont {B.}~\bibnamefont {Zhou}}, \bibinfo {author}
  {\bibfnamefont {Z.}~\bibnamefont {Wang}}, \bibinfo {author} {\bibfnamefont
  {Y.}~\bibnamefont {Zhang}}, \bibinfo {author} {\bibfnamefont
  {H.}~\bibnamefont {Weng}}, \bibinfo {author} {\bibfnamefont {D.}~\bibnamefont
  {Prabhakaran}}, \bibinfo {author} {\bibfnamefont {S.}~\bibnamefont {Mo}},
  \bibinfo {author} {\bibfnamefont {H.}~\bibnamefont {Peng}}, \bibinfo {author}
  {\bibfnamefont {P.}~\bibnamefont {Dudin}}, \emph {et~al.},\ }\bibfield
  {title} {\bibinfo {title} {A stable three-dimensional topological {Dirac}
  semimetal {Cd$_3$As$_2$}},\ }\href {https://doi.org/10.1038/nmat3990}
  {\bibfield  {journal} {\bibinfo  {journal} {Nat. Mater.}\ }\textbf {\bibinfo
  {volume} {13}},\ \bibinfo {pages} {677–681} (\bibinfo {year}
  {2014}{\natexlab{b}})}\BibitemShut {NoStop}%
\bibitem [{\citenamefont {Xu}\ \emph {et~al.}(2015{\natexlab{d}})\citenamefont
  {Xu}, \citenamefont {Liu}, \citenamefont {Kushwaha}, \citenamefont {Sankar},
  \citenamefont {Krizan}, \citenamefont {Belopolski}, \citenamefont {Neupane},
  \citenamefont {Bian}, \citenamefont {Alidoust}, \citenamefont {Chang} \emph
  {et~al.}}]{Xu-2015}%
  \BibitemOpen
  \bibfield  {author} {\bibinfo {author} {\bibfnamefont {S.-Y.}\ \bibnamefont
  {Xu}}, \bibinfo {author} {\bibfnamefont {C.}~\bibnamefont {Liu}}, \bibinfo
  {author} {\bibfnamefont {S.~K.}\ \bibnamefont {Kushwaha}}, \bibinfo {author}
  {\bibfnamefont {R.}~\bibnamefont {Sankar}}, \bibinfo {author} {\bibfnamefont
  {J.~W.}\ \bibnamefont {Krizan}}, \bibinfo {author} {\bibfnamefont
  {I.}~\bibnamefont {Belopolski}}, \bibinfo {author} {\bibfnamefont
  {M.}~\bibnamefont {Neupane}}, \bibinfo {author} {\bibfnamefont
  {G.}~\bibnamefont {Bian}}, \bibinfo {author} {\bibfnamefont {N.}~\bibnamefont
  {Alidoust}}, \bibinfo {author} {\bibfnamefont {T.-R.}\ \bibnamefont {Chang}},
  \emph {et~al.},\ }\bibfield  {title} {\bibinfo {title} {Observation of
  {{Fermi}} arc surface states in a topological metal},\ }\href
  {https://doi.org/10.1126/science.1256742} {\bibfield  {journal} {\bibinfo
  {journal} {Science}\ }\textbf {\bibinfo {volume} {347}},\ \bibinfo {pages}
  {294} (\bibinfo {year} {2015}{\natexlab{d}})}\BibitemShut {NoStop}%
\bibitem [{\citenamefont {Huang}\ \emph {et~al.}(2015)\citenamefont {Huang},
  \citenamefont {Xu}, \citenamefont {Belopolski}, \citenamefont {Lee},
  \citenamefont {Chang}, \citenamefont {Wang}, \citenamefont {Alidoust},
  \citenamefont {Bian}, \citenamefont {Neupane}, \citenamefont {Zhang} \emph
  {et~al.}}]{Huang-2015}%
  \BibitemOpen
  \bibfield  {author} {\bibinfo {author} {\bibfnamefont {S.-M.}\ \bibnamefont
  {Huang}}, \bibinfo {author} {\bibfnamefont {S.-Y.}\ \bibnamefont {Xu}},
  \bibinfo {author} {\bibfnamefont {I.}~\bibnamefont {Belopolski}}, \bibinfo
  {author} {\bibfnamefont {C.-C.}\ \bibnamefont {Lee}}, \bibinfo {author}
  {\bibfnamefont {G.}~\bibnamefont {Chang}}, \bibinfo {author} {\bibfnamefont
  {B.}~\bibnamefont {Wang}}, \bibinfo {author} {\bibfnamefont {N.}~\bibnamefont
  {Alidoust}}, \bibinfo {author} {\bibfnamefont {G.}~\bibnamefont {Bian}},
  \bibinfo {author} {\bibfnamefont {M.}~\bibnamefont {Neupane}}, \bibinfo
  {author} {\bibfnamefont {C.}~\bibnamefont {Zhang}}, \emph {et~al.},\
  }\bibfield  {title} {\bibinfo {title} {A {{Weyl Fermion}} semimetal with
  surface {{Fermi}} arcs in the transition metal monopnictide {{TaAs}} class},\
  }\href {https://doi.org/10.1038/ncomms8373} {\bibfield  {journal} {\bibinfo
  {journal} {Nat. Commun.}\ }\textbf {\bibinfo {volume} {6}},\ \bibinfo {pages}
  {7373} (\bibinfo {year} {2015})}\BibitemShut {NoStop}%
\bibitem [{\citenamefont {Weng}\ \emph {et~al.}(2015)\citenamefont {Weng},
  \citenamefont {Fang}, \citenamefont {Fang}, \citenamefont {Bernevig},\ and\
  \citenamefont {Dai}}]{wengWeylSemimetalPhase2015}%
  \BibitemOpen
  \bibfield  {author} {\bibinfo {author} {\bibfnamefont {H.}~\bibnamefont
  {Weng}}, \bibinfo {author} {\bibfnamefont {C.}~\bibnamefont {Fang}}, \bibinfo
  {author} {\bibfnamefont {Z.}~\bibnamefont {Fang}}, \bibinfo {author}
  {\bibfnamefont {B.~A.}\ \bibnamefont {Bernevig}},\ and\ \bibinfo {author}
  {\bibfnamefont {X.}~\bibnamefont {Dai}},\ }\bibfield  {title} {\bibinfo
  {title} {Weyl {{Semimetal Phase}} in {{Noncentrosymmetric Transition-Metal
  Monophosphides}}},\ }\href {https://doi.org/10.1103/PhysRevX.5.011029}
  {\bibfield  {journal} {\bibinfo  {journal} {Phys. Rev. X}\ }\textbf {\bibinfo
  {volume} {5}},\ \bibinfo {pages} {011029} (\bibinfo {year}
  {2015})}\BibitemShut {NoStop}%
\bibitem [{\citenamefont {Kim}\ \emph {et~al.}(2013)\citenamefont {Kim},
  \citenamefont {Kim}, \citenamefont {Wang}, \citenamefont {Sasaki},
  \citenamefont {Satoh}, \citenamefont {Ohnishi}, \citenamefont {Kitaura},
  \citenamefont {Yang},\ and\ \citenamefont {Li}}]{kimDiracWeylFermions2013}%
  \BibitemOpen
  \bibfield  {author} {\bibinfo {author} {\bibfnamefont {H.-J.}\ \bibnamefont
  {Kim}}, \bibinfo {author} {\bibfnamefont {K.-S.}\ \bibnamefont {Kim}},
  \bibinfo {author} {\bibfnamefont {J.-F.}\ \bibnamefont {Wang}}, \bibinfo
  {author} {\bibfnamefont {M.}~\bibnamefont {Sasaki}}, \bibinfo {author}
  {\bibfnamefont {N.}~\bibnamefont {Satoh}}, \bibinfo {author} {\bibfnamefont
  {A.}~\bibnamefont {Ohnishi}}, \bibinfo {author} {\bibfnamefont
  {M.}~\bibnamefont {Kitaura}}, \bibinfo {author} {\bibfnamefont
  {M.}~\bibnamefont {Yang}},\ and\ \bibinfo {author} {\bibfnamefont
  {L.}~\bibnamefont {Li}},\ }\bibfield  {title} {\bibinfo {title} {Dirac versus
  {{Weyl}} fermions in topological insulators: {{Adler-Bell-Jackiw}} anomaly in
  transport phenomena},\ }\href
  {https://doi.org/10.1103/PhysRevLett.111.246603} {\bibfield  {journal}
  {\bibinfo  {journal} {Phys. Rev. Lett.}\ }\textbf {\bibinfo {volume} {111}},\
  \bibinfo {pages} {246603} (\bibinfo {year} {2013})}\BibitemShut {NoStop}%
\bibitem [{\citenamefont {Liang}\ \emph {et~al.}(2015)\citenamefont {Liang},
  \citenamefont {Gibson}, \citenamefont {Ali}, \citenamefont {Liu},
  \citenamefont {Cava},\ and\ \citenamefont
  {Ong}}]{liangUltrahighMobilityGiant2015}%
  \BibitemOpen
  \bibfield  {author} {\bibinfo {author} {\bibfnamefont {T.}~\bibnamefont
  {Liang}}, \bibinfo {author} {\bibfnamefont {Q.}~\bibnamefont {Gibson}},
  \bibinfo {author} {\bibfnamefont {M.~N.}\ \bibnamefont {Ali}}, \bibinfo
  {author} {\bibfnamefont {M.}~\bibnamefont {Liu}}, \bibinfo {author}
  {\bibfnamefont {R.~J.}\ \bibnamefont {Cava}},\ and\ \bibinfo {author}
  {\bibfnamefont {N.~P.}\ \bibnamefont {Ong}},\ }\bibfield  {title} {\bibinfo
  {title} {Ultrahigh mobility and giant magnetoresistance in the {Dirac}
  semimetal {Cd$_3$As$_2$}},\ }\href {https://doi.org/10.1038/nmat4143}
  {\bibfield  {journal} {\bibinfo  {journal} {Nat. Mater.}\ }\textbf {\bibinfo
  {volume} {14}},\ \bibinfo {pages} {280} (\bibinfo {year} {2015})}\BibitemShut
  {NoStop}%
\bibitem [{\citenamefont {Li}\ \emph {et~al.}(2016)\citenamefont {Li},
  \citenamefont {Kharzeev}, \citenamefont {Zhang}, \citenamefont {Huang},
  \citenamefont {Pletikosić}, \citenamefont {Fedorov}, \citenamefont {Zhong},
  \citenamefont {Schneeloch}, \citenamefont {Gu},\ and\ \citenamefont
  {Valla}}]{liChiralMagneticEffect2016}%
  \BibitemOpen
  \bibfield  {author} {\bibinfo {author} {\bibfnamefont {Q.}~\bibnamefont
  {Li}}, \bibinfo {author} {\bibfnamefont {D.~E.}\ \bibnamefont {Kharzeev}},
  \bibinfo {author} {\bibfnamefont {C.}~\bibnamefont {Zhang}}, \bibinfo
  {author} {\bibfnamefont {Y.}~\bibnamefont {Huang}}, \bibinfo {author}
  {\bibfnamefont {I.}~\bibnamefont {Pletikosić}}, \bibinfo {author}
  {\bibfnamefont {A.~V.}\ \bibnamefont {Fedorov}}, \bibinfo {author}
  {\bibfnamefont {R.~D.}\ \bibnamefont {Zhong}}, \bibinfo {author}
  {\bibfnamefont {J.~A.}\ \bibnamefont {Schneeloch}}, \bibinfo {author}
  {\bibfnamefont {G.~D.}\ \bibnamefont {Gu}},\ and\ \bibinfo {author}
  {\bibfnamefont {T.}~\bibnamefont {Valla}},\ }\bibfield  {title} {\bibinfo
  {title} {Chiral magnetic effect in {ZrTe$_5$}},\ }\href
  {https://doi.org/10.1038/nphys3648} {\bibfield  {journal} {\bibinfo
  {journal} {Nat. Phys.}\ }\textbf {\bibinfo {volume} {12}},\ \bibinfo {pages}
  {550} (\bibinfo {year} {2016})}\BibitemShut {NoStop}%
\bibitem [{\citenamefont {Wang}\ \emph {et~al.}(2021)\citenamefont {Wang},
  \citenamefont {Cheng}, \citenamefont {Wang}, \citenamefont {Zhang},
  \citenamefont {Lu}, \citenamefont {Yi}, \citenamefont {Niu}, \citenamefont
  {Deng}, \citenamefont {Liu}, \citenamefont {Chen},\ and\ \citenamefont
  {Pan}}]{wang_realization_2021}%
  \BibitemOpen
  \bibfield  {author} {\bibinfo {author} {\bibfnamefont {Z.-Y.}\ \bibnamefont
  {Wang}}, \bibinfo {author} {\bibfnamefont {X.-C.}\ \bibnamefont {Cheng}},
  \bibinfo {author} {\bibfnamefont {B.-Z.}\ \bibnamefont {Wang}}, \bibinfo
  {author} {\bibfnamefont {J.-Y.}\ \bibnamefont {Zhang}}, \bibinfo {author}
  {\bibfnamefont {Y.-H.}\ \bibnamefont {Lu}}, \bibinfo {author} {\bibfnamefont
  {C.-R.}\ \bibnamefont {Yi}}, \bibinfo {author} {\bibfnamefont
  {S.}~\bibnamefont {Niu}}, \bibinfo {author} {\bibfnamefont {Y.}~\bibnamefont
  {Deng}}, \bibinfo {author} {\bibfnamefont {X.-J.}\ \bibnamefont {Liu}},
  \bibinfo {author} {\bibfnamefont {S.}~\bibnamefont {Chen}},\ and\ \bibinfo
  {author} {\bibfnamefont {J.-W.}\ \bibnamefont {Pan}},\ }\bibfield  {title}
  {\bibinfo {title} {Realization of an ideal {{Weyl}} semimetal band in a
  quantum gas with {{3D}} spin-orbit coupling},\ }\href
  {https://doi.org/10.1126/science.abc0105} {\bibfield  {journal} {\bibinfo
  {journal} {Science}\ }\textbf {\bibinfo {volume} {372}},\ \bibinfo {pages}
  {271} (\bibinfo {year} {2021})}\BibitemShut {NoStop}%
\bibitem [{\citenamefont {Syzranov}\ and\ \citenamefont
  {Radzihovsky}(2018)}]{syzranovHighdimensionalDisorderdrivenPhenomena2018}%
  \BibitemOpen
  \bibfield  {author} {\bibinfo {author} {\bibfnamefont {S.~V.}\ \bibnamefont
  {Syzranov}}\ and\ \bibinfo {author} {\bibfnamefont {L.}~\bibnamefont
  {Radzihovsky}},\ }\bibfield  {title} {\bibinfo {title} {High-dimensional
  disorder-driven phenomena in {{Weyl}} semimetals, semiconductors, and related
  systems},\ }\href {https://doi.org/10.1146/annurev-conmatphys-033117-054037}
  {\bibfield  {journal} {\bibinfo  {journal} {Annu. Rev. Condens. Matter
  Phys.}\ }\textbf {\bibinfo {volume} {9}},\ \bibinfo {pages} {35} (\bibinfo
  {year} {2018})}\BibitemShut {NoStop}%
\bibitem [{\citenamefont {Pixley}\ and\ \citenamefont
  {Wilson}(2021)}]{pixley_rare_2021}%
  \BibitemOpen
  \bibfield  {author} {\bibinfo {author} {\bibfnamefont {J.}~\bibnamefont
  {Pixley}}\ and\ \bibinfo {author} {\bibfnamefont {J.~H.}\ \bibnamefont
  {Wilson}},\ }\bibfield  {title} {\bibinfo {title} {Rare regions and avoided
  quantum criticality in disordered {{Weyl}} semimetals and superconductors},\
  }\href {https://doi.org/10.1016/j.aop.2021.168455} {\bibfield  {journal}
  {\bibinfo  {journal} {Ann. Phys. (N. Y.)}\ }\textbf {\bibinfo {volume}
  {435}},\ \bibinfo {pages} {168455} (\bibinfo {year} {2021})},\ \Eprint
  {https://arxiv.org/abs/2102.02822} {arxiv:2102.02822} \BibitemShut {NoStop}%
\bibitem [{\citenamefont {Shindou}\ and\ \citenamefont
  {Murakami}(2009)}]{shindou_effects_2009}%
  \BibitemOpen
  \bibfield  {author} {\bibinfo {author} {\bibfnamefont {R.}~\bibnamefont
  {Shindou}}\ and\ \bibinfo {author} {\bibfnamefont {S.}~\bibnamefont
  {Murakami}},\ }\bibfield  {title} {\bibinfo {title} {Effects of disorder in
  three-dimensional {{Z}} 2 quantum spin {{Hall}} systems},\ }\href
  {https://doi.org/10.1103/PhysRevB.79.045321} {\bibfield  {journal} {\bibinfo
  {journal} {Phys. Rev. B}\ }\textbf {\bibinfo {volume} {79}},\ \bibinfo
  {pages} {045321} (\bibinfo {year} {2009})}\BibitemShut {NoStop}%
\bibitem [{\citenamefont {Ominato}\ and\ \citenamefont
  {Koshino}(2014)}]{Ominato-2014}%
  \BibitemOpen
  \bibfield  {author} {\bibinfo {author} {\bibfnamefont {Y.}~\bibnamefont
  {Ominato}}\ and\ \bibinfo {author} {\bibfnamefont {M.}~\bibnamefont
  {Koshino}},\ }\bibfield  {title} {\bibinfo {title} {Quantum transport in a
  three-dimensional {{Weyl}} electron system},\ }\href
  {https://doi.org/10.1103/PhysRevB.89.054202} {\bibfield  {journal} {\bibinfo
  {journal} {Phys. Rev. B}\ }\textbf {\bibinfo {volume} {89}},\ \bibinfo
  {pages} {054202} (\bibinfo {year} {2014})}\BibitemShut {NoStop}%
\bibitem [{\citenamefont {Ryu}\ and\ \citenamefont
  {Nomura}(2012)}]{ryu_disorder-induced_2012}%
  \BibitemOpen
  \bibfield  {author} {\bibinfo {author} {\bibfnamefont {S.}~\bibnamefont
  {Ryu}}\ and\ \bibinfo {author} {\bibfnamefont {K.}~\bibnamefont {Nomura}},\
  }\bibfield  {title} {\bibinfo {title} {Disorder-induced quantum phase
  transitions in three-dimensional topological insulators and
  superconductors},\ }\href {https://doi.org/10.1103/PhysRevB.85.155138}
  {\bibfield  {journal} {\bibinfo  {journal} {Phys. Rev. B}\ }\textbf {\bibinfo
  {volume} {85}},\ \bibinfo {pages} {155138} (\bibinfo {year}
  {2012})}\BibitemShut {NoStop}%
\bibitem [{\citenamefont {Goswami}\ and\ \citenamefont
  {Chakravarty}(2011)}]{Goswami-2011}%
  \BibitemOpen
  \bibfield  {author} {\bibinfo {author} {\bibfnamefont {P.}~\bibnamefont
  {Goswami}}\ and\ \bibinfo {author} {\bibfnamefont {S.}~\bibnamefont
  {Chakravarty}},\ }\bibfield  {title} {\bibinfo {title} {Quantum criticality
  between topological and band insulators in 3+1 dimensions},\ }\href
  {https://doi.org/10.1103/PhysRevLett.107.196803} {\bibfield  {journal}
  {\bibinfo  {journal} {Phys. Rev. Lett.}\ }\textbf {\bibinfo {volume} {107}},\
  \bibinfo {pages} {196803} (\bibinfo {year} {2011})}\BibitemShut {NoStop}%
\bibitem [{\citenamefont {Syzranov}\ \emph {et~al.}(2015)\citenamefont
  {Syzranov}, \citenamefont {Radzihovsky},\ and\ \citenamefont
  {Gurarie}}]{Sergey-2015}%
  \BibitemOpen
  \bibfield  {author} {\bibinfo {author} {\bibfnamefont {S.~V.}\ \bibnamefont
  {Syzranov}}, \bibinfo {author} {\bibfnamefont {L.}~\bibnamefont
  {Radzihovsky}},\ and\ \bibinfo {author} {\bibfnamefont {V.}~\bibnamefont
  {Gurarie}},\ }\bibfield  {title} {\bibinfo {title} {Critical transport in
  weakly disordered semiconductors and semimetals},\ }\href
  {https://doi.org/10.1103/PhysRevLett.114.166601} {\bibfield  {journal}
  {\bibinfo  {journal} {Phys. Rev. Lett.}\ }\textbf {\bibinfo {volume} {114}},\
  \bibinfo {pages} {166601} (\bibinfo {year} {2015})}\BibitemShut {NoStop}%
\bibitem [{\citenamefont {Nandkishore}\ \emph {et~al.}(2014)\citenamefont
  {Nandkishore}, \citenamefont {Huse},\ and\ \citenamefont
  {Sondhi}}]{nandkishoreRareRegionEffects2014}%
  \BibitemOpen
  \bibfield  {author} {\bibinfo {author} {\bibfnamefont {R.}~\bibnamefont
  {Nandkishore}}, \bibinfo {author} {\bibfnamefont {D.~A.}\ \bibnamefont
  {Huse}},\ and\ \bibinfo {author} {\bibfnamefont {S.~L.}\ \bibnamefont
  {Sondhi}},\ }\bibfield  {title} {\bibinfo {title} {Rare region effects
  dominate weakly disordered three-dimensional {{Dirac}} points},\ }\href
  {https://doi.org/10.1103/PhysRevB.89.245110} {\bibfield  {journal} {\bibinfo
  {journal} {Phys. Rev. B}\ }\textbf {\bibinfo {volume} {89}},\ \bibinfo
  {pages} {245110} (\bibinfo {year} {2014})}\BibitemShut {NoStop}%
\bibitem [{\citenamefont {Pixley}\ \emph
  {et~al.}(2016{\natexlab{a}})\citenamefont {Pixley}, \citenamefont {Huse},\
  and\ \citenamefont {Das~Sarma}}]{Pixley-2016}%
  \BibitemOpen
  \bibfield  {author} {\bibinfo {author} {\bibfnamefont {J.~H.}\ \bibnamefont
  {Pixley}}, \bibinfo {author} {\bibfnamefont {D.~A.}\ \bibnamefont {Huse}},\
  and\ \bibinfo {author} {\bibfnamefont {S.}~\bibnamefont {Das~Sarma}},\
  }\bibfield  {title} {\bibinfo {title} {Rare-region-induced avoided quantum
  criticality in disordered three-dimensional {{Dirac}} and {{Weyl}}
  semimetals},\ }\href {https://doi.org/10.1103/PhysRevX.6.021042} {\bibfield
  {journal} {\bibinfo  {journal} {Phys. Rev. X}\ }\textbf {\bibinfo {volume}
  {6}},\ \bibinfo {pages} {021042} (\bibinfo {year}
  {2016}{\natexlab{a}})}\BibitemShut {NoStop}%
\bibitem [{\citenamefont {Pixley}\ \emph
  {et~al.}(2016{\natexlab{b}})\citenamefont {Pixley}, \citenamefont {Huse},\
  and\ \citenamefont {Das~Sarma}}]{PixleyBR-2016}%
  \BibitemOpen
  \bibfield  {author} {\bibinfo {author} {\bibfnamefont {J.~H.}\ \bibnamefont
  {Pixley}}, \bibinfo {author} {\bibfnamefont {D.~A.}\ \bibnamefont {Huse}},\
  and\ \bibinfo {author} {\bibfnamefont {S.}~\bibnamefont {Das~Sarma}},\
  }\bibfield  {title} {\bibinfo {title} {Uncovering the hidden quantum critical
  point in disordered massless dirac and weyl semimetals},\ }\href
  {https://doi.org/10.1103/PhysRevB.94.121107} {\bibfield  {journal} {\bibinfo
  {journal} {Phys. Rev. B}\ }\textbf {\bibinfo {volume} {94}},\ \bibinfo
  {pages} {121107} (\bibinfo {year} {2016}{\natexlab{b}})}\BibitemShut
  {NoStop}%
\bibitem [{\citenamefont {Pixley}\ \emph {et~al.}(2017)\citenamefont {Pixley},
  \citenamefont {Chou}, \citenamefont {Goswami}, \citenamefont {Huse},
  \citenamefont {Nandkishore}, \citenamefont {Radzihovsky},\ and\ \citenamefont
  {Das~Sarma}}]{pixleySingleparticleExcitationsDisordered2017}%
  \BibitemOpen
  \bibfield  {author} {\bibinfo {author} {\bibfnamefont {J.~H.}\ \bibnamefont
  {Pixley}}, \bibinfo {author} {\bibfnamefont {Y.-Z.}\ \bibnamefont {Chou}},
  \bibinfo {author} {\bibfnamefont {P.}~\bibnamefont {Goswami}}, \bibinfo
  {author} {\bibfnamefont {D.~A.}\ \bibnamefont {Huse}}, \bibinfo {author}
  {\bibfnamefont {R.}~\bibnamefont {Nandkishore}}, \bibinfo {author}
  {\bibfnamefont {L.}~\bibnamefont {Radzihovsky}},\ and\ \bibinfo {author}
  {\bibfnamefont {S.}~\bibnamefont {Das~Sarma}},\ }\bibfield  {title} {\bibinfo
  {title} {Single-particle excitations in disordered {{Weyl}} fluids},\ }\href
  {https://doi.org/10.1103/PhysRevB.95.235101} {\bibfield  {journal} {\bibinfo
  {journal} {Phys. Rev. B}\ }\textbf {\bibinfo {volume} {95}},\ \bibinfo
  {pages} {235101} (\bibinfo {year} {2017})}\BibitemShut {NoStop}%
\bibitem [{\citenamefont {Gurarie}(2017)}]{Guararie-2017}%
  \BibitemOpen
  \bibfield  {author} {\bibinfo {author} {\bibfnamefont {V.}~\bibnamefont
  {Gurarie}},\ }\bibfield  {title} {\bibinfo {title} {Theory of avoided
  criticality in quantum motion in a random potential in high dimensions},\
  }\href {https://doi.org/10.1103/PhysRevB.96.014205} {\bibfield  {journal}
  {\bibinfo  {journal} {Phys. Rev. B}\ }\textbf {\bibinfo {volume} {96}},\
  \bibinfo {pages} {014205} (\bibinfo {year} {2017})}\BibitemShut {NoStop}%
\bibitem [{\citenamefont {Buchhold}\ \emph
  {et~al.}(2018{\natexlab{a}})\citenamefont {Buchhold}, \citenamefont {Diehl},\
  and\ \citenamefont {Altland}}]{Buchold-2018}%
  \BibitemOpen
  \bibfield  {author} {\bibinfo {author} {\bibfnamefont {M.}~\bibnamefont
  {Buchhold}}, \bibinfo {author} {\bibfnamefont {S.}~\bibnamefont {Diehl}},\
  and\ \bibinfo {author} {\bibfnamefont {A.}~\bibnamefont {Altland}},\
  }\bibfield  {title} {\bibinfo {title} {Vanishing density of states in weakly
  disordered weyl semimetals},\ }\href
  {https://doi.org/10.1103/PhysRevLett.121.215301} {\bibfield  {journal}
  {\bibinfo  {journal} {Phys. Rev. Lett.}\ }\textbf {\bibinfo {volume} {121}},\
  \bibinfo {pages} {215301} (\bibinfo {year} {2018}{\natexlab{a}})}\BibitemShut
  {NoStop}%
\bibitem [{\citenamefont {Buchhold}\ \emph
  {et~al.}(2018{\natexlab{b}})\citenamefont {Buchhold}, \citenamefont {Diehl},\
  and\ \citenamefont {Altland}}]{Buchold-B2018}%
  \BibitemOpen
  \bibfield  {author} {\bibinfo {author} {\bibfnamefont {M.}~\bibnamefont
  {Buchhold}}, \bibinfo {author} {\bibfnamefont {S.}~\bibnamefont {Diehl}},\
  and\ \bibinfo {author} {\bibfnamefont {A.}~\bibnamefont {Altland}},\
  }\bibfield  {title} {\bibinfo {title} {Nodal points of weyl semimetals
  survive the presence of moderate disorder},\ }\href
  {https://doi.org/10.1103/PhysRevB.98.205134} {\bibfield  {journal} {\bibinfo
  {journal} {Phys. Rev. B}\ }\textbf {\bibinfo {volume} {98}},\ \bibinfo
  {pages} {205134} (\bibinfo {year} {2018}{\natexlab{b}})}\BibitemShut
  {NoStop}%
\bibitem [{\citenamefont {Galindo}\ and\ \citenamefont
  {Pascual}(2012)}]{galindo2012quantum}%
  \BibitemOpen
  \bibfield  {author} {\bibinfo {author} {\bibfnamefont {A.}~\bibnamefont
  {Galindo}}\ and\ \bibinfo {author} {\bibfnamefont {P.}~\bibnamefont
  {Pascual}},\ }\href@noop {} {\emph {\bibinfo {title} {Quantum mechanics I}}}\
  (\bibinfo  {publisher} {Springer Science \& Business Media},\ \bibinfo {year}
  {2012})\BibitemShut {NoStop}%
\bibitem [{\citenamefont {Pires}\ \emph {et~al.}(2021)\citenamefont {Pires},
  \citenamefont {Amorim}, \citenamefont {Ferreira}, \citenamefont {Adagideli},
  \citenamefont {Mucciolo},\ and\ \citenamefont {Lopes}}]{Pires-2021}%
  \BibitemOpen
  \bibfield  {author} {\bibinfo {author} {\bibfnamefont {J.~P.~S.}\
  \bibnamefont {Pires}}, \bibinfo {author} {\bibfnamefont {B.}~\bibnamefont
  {Amorim}}, \bibinfo {author} {\bibfnamefont {A.}~\bibnamefont {Ferreira}},
  \bibinfo {author} {\bibfnamefont {i.~d. I. m.~c.}\ \bibnamefont {Adagideli}},
  \bibinfo {author} {\bibfnamefont {E.~R.}\ \bibnamefont {Mucciolo}},\ and\
  \bibinfo {author} {\bibfnamefont {J.~M. V.~P.}\ \bibnamefont {Lopes}},\
  }\bibfield  {title} {\bibinfo {title} {Breakdown of universality in
  three-dimensional dirac semimetals with random impurities},\ }\href
  {https://doi.org/10.1103/PhysRevResearch.3.013183} {\bibfield  {journal}
  {\bibinfo  {journal} {Phys. Rev. Res.}\ }\textbf {\bibinfo {volume} {3}},\
  \bibinfo {pages} {013183} (\bibinfo {year} {2021})}\BibitemShut {NoStop}%
\bibitem [{\citenamefont {Wilson}\ \emph
  {et~al.}(2020{\natexlab{a}})\citenamefont {Wilson}, \citenamefont {Huse},
  \citenamefont {Das~Sarma},\ and\ \citenamefont
  {Pixley}}]{wilsonAvoidedQuantumCriticality2020}%
  \BibitemOpen
  \bibfield  {author} {\bibinfo {author} {\bibfnamefont {J.~H.}\ \bibnamefont
  {Wilson}}, \bibinfo {author} {\bibfnamefont {D.~A.}\ \bibnamefont {Huse}},
  \bibinfo {author} {\bibfnamefont {S.}~\bibnamefont {Das~Sarma}},\ and\
  \bibinfo {author} {\bibfnamefont {J.~H.}\ \bibnamefont {Pixley}},\ }\bibfield
   {title} {\bibinfo {title} {Avoided quantum criticality in exact numerical
  simulations of a single disordered {{Weyl}} cone},\ }\href
  {https://doi.org/10.1103/PhysRevB.102.100201} {\bibfield  {journal} {\bibinfo
   {journal} {Phys. Rev. B}\ }\textbf {\bibinfo {volume} {102}},\ \bibinfo
  {pages} {100201} (\bibinfo {year} {2020}{\natexlab{a}})}\BibitemShut
  {NoStop}%
\bibitem [{\citenamefont {Pixley}\ \emph {et~al.}(2018)\citenamefont {Pixley},
  \citenamefont {Wilson}, \citenamefont {Huse},\ and\ \citenamefont
  {Gopalakrishnan}}]{Pixley-2018}%
  \BibitemOpen
  \bibfield  {author} {\bibinfo {author} {\bibfnamefont {J.~H.}\ \bibnamefont
  {Pixley}}, \bibinfo {author} {\bibfnamefont {J.~H.}\ \bibnamefont {Wilson}},
  \bibinfo {author} {\bibfnamefont {D.~A.}\ \bibnamefont {Huse}},\ and\
  \bibinfo {author} {\bibfnamefont {S.}~\bibnamefont {Gopalakrishnan}},\
  }\bibfield  {title} {\bibinfo {title} {Weyl semimetal to metal phase
  transitions driven by quasiperiodic potentials},\ }\href
  {https://doi.org/10.1103/PhysRevLett.120.207604} {\bibfield  {journal}
  {\bibinfo  {journal} {Phys. Rev. Lett.}\ }\textbf {\bibinfo {volume} {120}},\
  \bibinfo {pages} {207604} (\bibinfo {year} {2018})}\BibitemShut {NoStop}%
\bibitem [{\citenamefont
  {Mastropietro}(2020)}]{mastropietroStabilityWeylSemimetals2020}%
  \BibitemOpen
  \bibfield  {author} {\bibinfo {author} {\bibfnamefont {V.}~\bibnamefont
  {Mastropietro}},\ }\bibfield  {title} {\bibinfo {title} {Stability of
  {{Weyl}} semimetals with quasiperiodic disorder},\ }\href
  {https://doi.org/10.1103/PhysRevB.102.045101} {\bibfield  {journal} {\bibinfo
   {journal} {Phys. Rev. B}\ }\textbf {\bibinfo {volume} {102}},\ \bibinfo
  {pages} {045101} (\bibinfo {year} {2020})}\BibitemShut {NoStop}%
\bibitem [{\citenamefont {Fu}\ \emph {et~al.}(2020)\citenamefont {Fu},
  \citenamefont {K{\"o}nig}, \citenamefont {Wilson}, \citenamefont {Chou},\
  and\ \citenamefont {Pixley}}]{fuMagicangleSemimetals2020}%
  \BibitemOpen
  \bibfield  {author} {\bibinfo {author} {\bibfnamefont {Y.}~\bibnamefont
  {Fu}}, \bibinfo {author} {\bibfnamefont {E.~J.}\ \bibnamefont {K{\"o}nig}},
  \bibinfo {author} {\bibfnamefont {J.~H.}\ \bibnamefont {Wilson}}, \bibinfo
  {author} {\bibfnamefont {Y.-Z.}\ \bibnamefont {Chou}},\ and\ \bibinfo
  {author} {\bibfnamefont {J.~H.}\ \bibnamefont {Pixley}},\ }\bibfield  {title}
  {\bibinfo {title} {Magic-angle semimetals},\ }\href
  {https://doi.org/10.1038/s41535-020-00271-9} {\bibfield  {journal} {\bibinfo
  {journal} {npj Quantum Mater.}\ }\textbf {\bibinfo {volume} {5}},\ \bibinfo
  {pages} {1} (\bibinfo {year} {2020})}\BibitemShut {NoStop}%
\bibitem [{\citenamefont {Chou}\ \emph {et~al.}(2020)\citenamefont {Chou},
  \citenamefont {Fu}, \citenamefont {Wilson}, \citenamefont {K{\"o}nig},\ and\
  \citenamefont {Pixley}}]{chou_magic-angle_2020}%
  \BibitemOpen
  \bibfield  {author} {\bibinfo {author} {\bibfnamefont {Y.-Z.}\ \bibnamefont
  {Chou}}, \bibinfo {author} {\bibfnamefont {Y.}~\bibnamefont {Fu}}, \bibinfo
  {author} {\bibfnamefont {J.~H.}\ \bibnamefont {Wilson}}, \bibinfo {author}
  {\bibfnamefont {E.~J.}\ \bibnamefont {K{\"o}nig}},\ and\ \bibinfo {author}
  {\bibfnamefont {J.~H.}\ \bibnamefont {Pixley}},\ }\bibfield  {title}
  {\bibinfo {title} {Magic-angle semimetals with chiral symmetry},\ }\href
  {https://doi.org/10.1103/PhysRevB.101.235121} {\bibfield  {journal} {\bibinfo
   {journal} {Phys. Rev. B}\ }\textbf {\bibinfo {volume} {101}},\ \bibinfo
  {pages} {235121} (\bibinfo {year} {2020})}\BibitemShut {NoStop}%
\bibitem [{\citenamefont {Fu}\ \emph {et~al.}(2021)\citenamefont {Fu},
  \citenamefont {Wilson},\ and\ \citenamefont {Pixley}}]{Fu-2021}%
  \BibitemOpen
  \bibfield  {author} {\bibinfo {author} {\bibfnamefont {Y.}~\bibnamefont
  {Fu}}, \bibinfo {author} {\bibfnamefont {J.~H.}\ \bibnamefont {Wilson}},\
  and\ \bibinfo {author} {\bibfnamefont {J.~H.}\ \bibnamefont {Pixley}},\
  }\bibfield  {title} {\bibinfo {title} {Flat topological bands and eigenstate
  criticality in a quasiperiodic insulator},\ }\href
  {https://doi.org/10.1103/PhysRevB.104.L041106} {\bibfield  {journal}
  {\bibinfo  {journal} {Phys. Rev. B}\ }\textbf {\bibinfo {volume} {104}},\
  \bibinfo {pages} {L041106} (\bibinfo {year} {2021})}\BibitemShut {NoStop}%
\bibitem [{\citenamefont {Bistritzer}\ and\ \citenamefont
  {MacDonald}(2011)}]{bistritzer2011moire}%
  \BibitemOpen
  \bibfield  {author} {\bibinfo {author} {\bibfnamefont {R.}~\bibnamefont
  {Bistritzer}}\ and\ \bibinfo {author} {\bibfnamefont {A.~H.}\ \bibnamefont
  {MacDonald}},\ }\bibfield  {title} {\bibinfo {title} {Moir{\'e} bands in
  twisted double-layer graphene},\ }\href@noop {} {\bibfield  {journal}
  {\bibinfo  {journal} {Proceedings of the National Academy of Sciences}\
  }\textbf {\bibinfo {volume} {108}},\ \bibinfo {pages} {12233} (\bibinfo
  {year} {2011})}\BibitemShut {NoStop}%
\bibitem [{\citenamefont {Gon{\c{c}}alves}\ \emph {et~al.}(2021)\citenamefont
  {Gon{\c{c}}alves}, \citenamefont {Olyaei}, \citenamefont {Amorim},
  \citenamefont {Mondaini}, \citenamefont {Ribeiro},\ and\ \citenamefont
  {Castro}}]{gonccalves2021incommensurability}%
  \BibitemOpen
  \bibfield  {author} {\bibinfo {author} {\bibfnamefont {M.}~\bibnamefont
  {Gon{\c{c}}alves}}, \bibinfo {author} {\bibfnamefont {H.~Z.}\ \bibnamefont
  {Olyaei}}, \bibinfo {author} {\bibfnamefont {B.}~\bibnamefont {Amorim}},
  \bibinfo {author} {\bibfnamefont {R.}~\bibnamefont {Mondaini}}, \bibinfo
  {author} {\bibfnamefont {P.}~\bibnamefont {Ribeiro}},\ and\ \bibinfo {author}
  {\bibfnamefont {E.~V.}\ \bibnamefont {Castro}},\ }\bibfield  {title}
  {\bibinfo {title} {Incommensurability-induced sub-ballistic
  narrow-band-states in twisted bilayer graphene},\ }\href@noop {} {\bibfield
  {journal} {\bibinfo  {journal} {2D Materials}\ }\textbf {\bibinfo {volume}
  {9}},\ \bibinfo {pages} {011001} (\bibinfo {year} {2021})}\BibitemShut
  {NoStop}%
\bibitem [{\citenamefont {Beechem}\ \emph {et~al.}(2014)\citenamefont
  {Beechem}, \citenamefont {Ohta}, \citenamefont {Diaconescu},\ and\
  \citenamefont {Robinson}}]{beechem_rotational_2014}%
  \BibitemOpen
  \bibfield  {author} {\bibinfo {author} {\bibfnamefont {T.~E.}\ \bibnamefont
  {Beechem}}, \bibinfo {author} {\bibfnamefont {T.}~\bibnamefont {Ohta}},
  \bibinfo {author} {\bibfnamefont {B.}~\bibnamefont {Diaconescu}},\ and\
  \bibinfo {author} {\bibfnamefont {J.~T.}\ \bibnamefont {Robinson}},\
  }\bibfield  {title} {\bibinfo {title} {Rotational {{Disorder}} in {{Twisted
  Bilayer Graphene}}},\ }\href {https://doi.org/10.1021/nn405999z} {\bibfield
  {journal} {\bibinfo  {journal} {ACS Nano}\ }\textbf {\bibinfo {volume} {8}},\
  \bibinfo {pages} {1655} (\bibinfo {year} {2014})}\BibitemShut {NoStop}%
\bibitem [{\citenamefont {Wilson}\ \emph
  {et~al.}(2020{\natexlab{b}})\citenamefont {Wilson}, \citenamefont {Fu},
  \citenamefont {Das~Sarma},\ and\ \citenamefont
  {Pixley}}]{wilson_disorder_2020}%
  \BibitemOpen
  \bibfield  {author} {\bibinfo {author} {\bibfnamefont {J.~H.}\ \bibnamefont
  {Wilson}}, \bibinfo {author} {\bibfnamefont {Y.}~\bibnamefont {Fu}}, \bibinfo
  {author} {\bibfnamefont {S.}~\bibnamefont {Das~Sarma}},\ and\ \bibinfo
  {author} {\bibfnamefont {J.~H.}\ \bibnamefont {Pixley}},\ }\bibfield  {title}
  {\bibinfo {title} {Disorder in twisted bilayer graphene},\ }\href
  {https://doi.org/10.1103/PhysRevResearch.2.023325} {\bibfield  {journal}
  {\bibinfo  {journal} {Phys. Rev. Research}\ }\textbf {\bibinfo {volume}
  {2}},\ \bibinfo {pages} {023325} (\bibinfo {year}
  {2020}{\natexlab{b}})}\BibitemShut {NoStop}%
\bibitem [{\citenamefont {Uri}\ \emph {et~al.}(2020)\citenamefont {Uri},
  \citenamefont {Grover}, \citenamefont {Cao}, \citenamefont {Crosse},
  \citenamefont {Bagani}, \citenamefont {Rodan-Legrain}, \citenamefont
  {Myasoedov}, \citenamefont {Watanabe}, \citenamefont {Taniguchi},
  \citenamefont {Moon} \emph {et~al.}}]{uri2020mapping}%
  \BibitemOpen
  \bibfield  {author} {\bibinfo {author} {\bibfnamefont {A.}~\bibnamefont
  {Uri}}, \bibinfo {author} {\bibfnamefont {S.}~\bibnamefont {Grover}},
  \bibinfo {author} {\bibfnamefont {Y.}~\bibnamefont {Cao}}, \bibinfo {author}
  {\bibfnamefont {J.~A.}\ \bibnamefont {Crosse}}, \bibinfo {author}
  {\bibfnamefont {K.}~\bibnamefont {Bagani}}, \bibinfo {author} {\bibfnamefont
  {D.}~\bibnamefont {Rodan-Legrain}}, \bibinfo {author} {\bibfnamefont
  {Y.}~\bibnamefont {Myasoedov}}, \bibinfo {author} {\bibfnamefont
  {K.}~\bibnamefont {Watanabe}}, \bibinfo {author} {\bibfnamefont
  {T.}~\bibnamefont {Taniguchi}}, \bibinfo {author} {\bibfnamefont
  {P.}~\bibnamefont {Moon}}, \emph {et~al.},\ }\bibfield  {title} {\bibinfo
  {title} {Mapping the twist-angle disorder and landau levels in magic-angle
  graphene},\ }\href@noop {} {\bibfield  {journal} {\bibinfo  {journal}
  {Nature}\ }\textbf {\bibinfo {volume} {581}},\ \bibinfo {pages} {47}
  (\bibinfo {year} {2020})}\BibitemShut {NoStop}%
\bibitem [{\citenamefont {Padhi}\ \emph {et~al.}(2020)\citenamefont {Padhi},
  \citenamefont {Tiwari}, \citenamefont {Neupert},\ and\ \citenamefont
  {Ryu}}]{PhysRevResearch.2.033458}%
  \BibitemOpen
  \bibfield  {author} {\bibinfo {author} {\bibfnamefont {B.}~\bibnamefont
  {Padhi}}, \bibinfo {author} {\bibfnamefont {A.}~\bibnamefont {Tiwari}},
  \bibinfo {author} {\bibfnamefont {T.}~\bibnamefont {Neupert}},\ and\ \bibinfo
  {author} {\bibfnamefont {S.}~\bibnamefont {Ryu}},\ }\bibfield  {title}
  {\bibinfo {title} {Transport across twist angle domains in moir\'e
  graphene},\ }\href {https://doi.org/10.1103/PhysRevResearch.2.033458}
  {\bibfield  {journal} {\bibinfo  {journal} {Phys. Rev. Res.}\ }\textbf
  {\bibinfo {volume} {2}},\ \bibinfo {pages} {033458} (\bibinfo {year}
  {2020})}\BibitemShut {NoStop}%
\bibitem [{\citenamefont {Joy}\ \emph {et~al.}(2020)\citenamefont {Joy},
  \citenamefont {Khalid},\ and\ \citenamefont
  {Skinner}}]{PhysRevResearch.2.043416}%
  \BibitemOpen
  \bibfield  {author} {\bibinfo {author} {\bibfnamefont {S.}~\bibnamefont
  {Joy}}, \bibinfo {author} {\bibfnamefont {S.}~\bibnamefont {Khalid}},\ and\
  \bibinfo {author} {\bibfnamefont {B.}~\bibnamefont {Skinner}},\ }\bibfield
  {title} {\bibinfo {title} {Transparent mirror effect in
  twist-angle-disordered bilayer graphene},\ }\href
  {https://doi.org/10.1103/PhysRevResearch.2.043416} {\bibfield  {journal}
  {\bibinfo  {journal} {Phys. Rev. Res.}\ }\textbf {\bibinfo {volume} {2}},\
  \bibinfo {pages} {043416} (\bibinfo {year} {2020})}\BibitemShut {NoStop}%
\bibitem [{\citenamefont {Pixley}\ \emph
  {et~al.}(2016{\natexlab{c}})\citenamefont {Pixley}, \citenamefont {Huse},\
  and\ \citenamefont {Das~Sarma}}]{pixleyUncoveringHiddenQuantum2016}%
  \BibitemOpen
  \bibfield  {author} {\bibinfo {author} {\bibfnamefont {J.~H.}\ \bibnamefont
  {Pixley}}, \bibinfo {author} {\bibfnamefont {D.~A.}\ \bibnamefont {Huse}},\
  and\ \bibinfo {author} {\bibfnamefont {S.}~\bibnamefont {Das~Sarma}},\
  }\bibfield  {title} {\bibinfo {title} {Uncovering the hidden quantum critical
  point in disordered massless {{Dirac}} and {{Weyl}} semimetals},\ }\href
  {https://doi.org/10.1103/PhysRevB.94.121107} {\bibfield  {journal} {\bibinfo
  {journal} {Phys. Rev. B}\ }\textbf {\bibinfo {volume} {94}},\ \bibinfo
  {pages} {121107} (\bibinfo {year} {2016}{\natexlab{c}})}\BibitemShut
  {NoStop}%
\bibitem [{\citenamefont {Marzari}\ \emph {et~al.}(2012)\citenamefont
  {Marzari}, \citenamefont {Mostofi}, \citenamefont {Yates}, \citenamefont
  {Souza},\ and\ \citenamefont {Vanderbilt}}]{RevModPhys.84.1419}%
  \BibitemOpen
  \bibfield  {author} {\bibinfo {author} {\bibfnamefont {N.}~\bibnamefont
  {Marzari}}, \bibinfo {author} {\bibfnamefont {A.~A.}\ \bibnamefont
  {Mostofi}}, \bibinfo {author} {\bibfnamefont {J.~R.}\ \bibnamefont {Yates}},
  \bibinfo {author} {\bibfnamefont {I.}~\bibnamefont {Souza}},\ and\ \bibinfo
  {author} {\bibfnamefont {D.}~\bibnamefont {Vanderbilt}},\ }\bibfield  {title}
  {\bibinfo {title} {Maximally localized wannier functions: Theory and
  applications},\ }\href {https://doi.org/10.1103/RevModPhys.84.1419}
  {\bibfield  {journal} {\bibinfo  {journal} {Rev. Mod. Phys.}\ }\textbf
  {\bibinfo {volume} {84}},\ \bibinfo {pages} {1419} (\bibinfo {year}
  {2012})}\BibitemShut {NoStop}%
\bibitem [{\citenamefont {Goodman}(2007)}]{goodman2007speckle}%
  \BibitemOpen
  \bibfield  {author} {\bibinfo {author} {\bibfnamefont {J.~W.}\ \bibnamefont
  {Goodman}},\ }\href@noop {} {\emph {\bibinfo {title} {Speckle phenomena in
  optics: theory and applications}}}\ (\bibinfo  {publisher} {Roberts and
  Company Publishers},\ \bibinfo {year} {2007})\BibitemShut {NoStop}%
\bibitem [{\citenamefont {Monroe}\ \emph {et~al.}(2021)\citenamefont {Monroe},
  \citenamefont {Campbell}, \citenamefont {Duan}, \citenamefont {Gong},
  \citenamefont {Gorshkov}, \citenamefont {Hess}, \citenamefont {Islam},
  \citenamefont {Kim}, \citenamefont {Linke}, \citenamefont {Pagano},
  \citenamefont {Richerme}, \citenamefont {Senko},\ and\ \citenamefont
  {Yao}}]{RevModPhys.93.025001}%
  \BibitemOpen
  \bibfield  {author} {\bibinfo {author} {\bibfnamefont {C.}~\bibnamefont
  {Monroe}}, \bibinfo {author} {\bibfnamefont {W.~C.}\ \bibnamefont
  {Campbell}}, \bibinfo {author} {\bibfnamefont {L.-M.}\ \bibnamefont {Duan}},
  \bibinfo {author} {\bibfnamefont {Z.-X.}\ \bibnamefont {Gong}}, \bibinfo
  {author} {\bibfnamefont {A.~V.}\ \bibnamefont {Gorshkov}}, \bibinfo {author}
  {\bibfnamefont {P.~W.}\ \bibnamefont {Hess}}, \bibinfo {author}
  {\bibfnamefont {R.}~\bibnamefont {Islam}}, \bibinfo {author} {\bibfnamefont
  {K.}~\bibnamefont {Kim}}, \bibinfo {author} {\bibfnamefont {N.~M.}\
  \bibnamefont {Linke}}, \bibinfo {author} {\bibfnamefont {G.}~\bibnamefont
  {Pagano}}, \bibinfo {author} {\bibfnamefont {P.}~\bibnamefont {Richerme}},
  \bibinfo {author} {\bibfnamefont {C.}~\bibnamefont {Senko}},\ and\ \bibinfo
  {author} {\bibfnamefont {N.~Y.}\ \bibnamefont {Yao}},\ }\bibfield  {title}
  {\bibinfo {title} {Programmable quantum simulations of spin systems with
  trapped ions},\ }\href {https://doi.org/10.1103/RevModPhys.93.025001}
  {\bibfield  {journal} {\bibinfo  {journal} {Rev. Mod. Phys.}\ }\textbf
  {\bibinfo {volume} {93}},\ \bibinfo {pages} {025001} (\bibinfo {year}
  {2021})}\BibitemShut {NoStop}%
\bibitem [{\citenamefont {An}\ \emph {et~al.}(2021)\citenamefont {An},
  \citenamefont {Padavi\ifmmode~\acute{c}\else \'{c}\fi{}}, \citenamefont
  {Meier}, \citenamefont {Hegde}, \citenamefont {Ganeshan}, \citenamefont
  {Pixley}, \citenamefont {Vishveshwara},\ and\ \citenamefont
  {Gadway}}]{PhysRevLett.126.040603}%
  \BibitemOpen
  \bibfield  {author} {\bibinfo {author} {\bibfnamefont {F.~A.}\ \bibnamefont
  {An}}, \bibinfo {author} {\bibfnamefont {K.}~\bibnamefont
  {Padavi\ifmmode~\acute{c}\else \'{c}\fi{}}}, \bibinfo {author} {\bibfnamefont
  {E.~J.}\ \bibnamefont {Meier}}, \bibinfo {author} {\bibfnamefont
  {S.}~\bibnamefont {Hegde}}, \bibinfo {author} {\bibfnamefont
  {S.}~\bibnamefont {Ganeshan}}, \bibinfo {author} {\bibfnamefont {J.~H.}\
  \bibnamefont {Pixley}}, \bibinfo {author} {\bibfnamefont {S.}~\bibnamefont
  {Vishveshwara}},\ and\ \bibinfo {author} {\bibfnamefont {B.}~\bibnamefont
  {Gadway}},\ }\bibfield  {title} {\bibinfo {title} {Interactions and mobility
  edges: Observing the generalized aubry-andr\'e model},\ }\href
  {https://doi.org/10.1103/PhysRevLett.126.040603} {\bibfield  {journal}
  {\bibinfo  {journal} {Phys. Rev. Lett.}\ }\textbf {\bibinfo {volume} {126}},\
  \bibinfo {pages} {040603} (\bibinfo {year} {2021})}\BibitemShut {NoStop}%
\bibitem [{\citenamefont {Stuart}\ and\ \citenamefont
  {Kuhn}(2018)}]{stuart2018single}%
  \BibitemOpen
  \bibfield  {author} {\bibinfo {author} {\bibfnamefont {D.}~\bibnamefont
  {Stuart}}\ and\ \bibinfo {author} {\bibfnamefont {A.}~\bibnamefont {Kuhn}},\
  }\bibfield  {title} {\bibinfo {title} {Single-atom trapping and transport in
  dmd-controlled optical tweezers},\ }\href@noop {} {\bibfield  {journal}
  {\bibinfo  {journal} {New Journal of Physics}\ }\textbf {\bibinfo {volume}
  {20}},\ \bibinfo {pages} {023013} (\bibinfo {year} {2018})}\BibitemShut
  {NoStop}%
\bibitem [{\citenamefont {Lett}\ \emph {et~al.}(1988)\citenamefont {Lett},
  \citenamefont {Watts}, \citenamefont {Westbrook}, \citenamefont {Phillips},
  \citenamefont {Gould},\ and\ \citenamefont
  {Metcalf}}]{lett_observation_1988}%
  \BibitemOpen
  \bibfield  {author} {\bibinfo {author} {\bibfnamefont {P.~D.}\ \bibnamefont
  {Lett}}, \bibinfo {author} {\bibfnamefont {R.~N.}\ \bibnamefont {Watts}},
  \bibinfo {author} {\bibfnamefont {C.~I.}\ \bibnamefont {Westbrook}}, \bibinfo
  {author} {\bibfnamefont {W.~D.}\ \bibnamefont {Phillips}}, \bibinfo {author}
  {\bibfnamefont {P.~L.}\ \bibnamefont {Gould}},\ and\ \bibinfo {author}
  {\bibfnamefont {H.~J.}\ \bibnamefont {Metcalf}},\ }\bibfield  {title}
  {\bibinfo {title} {Observation of {{Atoms Laser Cooled}} below the {{Doppler
  Limit}}},\ }\href {https://doi.org/10.1103/PhysRevLett.61.169} {\bibfield
  {journal} {\bibinfo  {journal} {Phys. Rev. Lett.}\ }\textbf {\bibinfo
  {volume} {61}},\ \bibinfo {pages} {169} (\bibinfo {year} {1988})}\BibitemShut
  {NoStop}%
\bibitem [{\citenamefont {Gupta}\ \emph {et~al.}(2003)\citenamefont {Gupta},
  \citenamefont {Hadzibabic}, \citenamefont {Zwierlein}, \citenamefont {Stan},
  \citenamefont {Dieckmann}, \citenamefont {Schunck}, \citenamefont {{van
  Kempen}}, \citenamefont {Verhaar},\ and\ \citenamefont
  {Ketterle}}]{gupta_radio-frequency_2003}%
  \BibitemOpen
  \bibfield  {author} {\bibinfo {author} {\bibfnamefont {S.}~\bibnamefont
  {Gupta}}, \bibinfo {author} {\bibfnamefont {Z.}~\bibnamefont {Hadzibabic}},
  \bibinfo {author} {\bibfnamefont {M.~W.}\ \bibnamefont {Zwierlein}}, \bibinfo
  {author} {\bibfnamefont {C.~A.}\ \bibnamefont {Stan}}, \bibinfo {author}
  {\bibfnamefont {K.}~\bibnamefont {Dieckmann}}, \bibinfo {author}
  {\bibfnamefont {C.~H.}\ \bibnamefont {Schunck}}, \bibinfo {author}
  {\bibfnamefont {E.~G.~M.}\ \bibnamefont {{van Kempen}}}, \bibinfo {author}
  {\bibfnamefont {B.~J.}\ \bibnamefont {Verhaar}},\ and\ \bibinfo {author}
  {\bibfnamefont {W.}~\bibnamefont {Ketterle}},\ }\bibfield  {title} {\bibinfo
  {title} {Radio-{{Frequency Spectroscopy}} of {{Ultracold Fermions}}},\ }\href
  {https://doi.org/10.1126/science.1085335} {\bibfield  {journal} {\bibinfo
  {journal} {Science}\ }\textbf {\bibinfo {volume} {300}},\ \bibinfo {pages}
  {1723} (\bibinfo {year} {2003})}\BibitemShut {NoStop}%
\bibitem [{\citenamefont {Yi}\ \emph {et~al.}(2022)\citenamefont {Yi},
  \citenamefont {K\"onig},\ and\ \citenamefont {Pixley}}]{PhysRevB.106.195123}%
  \BibitemOpen
  \bibfield  {author} {\bibinfo {author} {\bibfnamefont {J.}~\bibnamefont
  {Yi}}, \bibinfo {author} {\bibfnamefont {E.~J.}\ \bibnamefont {K\"onig}},\
  and\ \bibinfo {author} {\bibfnamefont {J.~H.}\ \bibnamefont {Pixley}},\
  }\bibfield  {title} {\bibinfo {title} {Low energy excitation spectrum of
  magic-angle semimetals},\ }\href
  {https://doi.org/10.1103/PhysRevB.106.195123} {\bibfield  {journal} {\bibinfo
   {journal} {Phys. Rev. B}\ }\textbf {\bibinfo {volume} {106}},\ \bibinfo
  {pages} {195123} (\bibinfo {year} {2022})}\BibitemShut {NoStop}%
\bibitem [{\citenamefont {Herrig}\ \emph {et~al.}(2022)\citenamefont {Herrig},
  \citenamefont {Pixley}, \citenamefont {K{\"o}nig},\ and\ \citenamefont
  {Riwar}}]{herrig2022quasiperiodic}%
  \BibitemOpen
  \bibfield  {author} {\bibinfo {author} {\bibfnamefont {T.}~\bibnamefont
  {Herrig}}, \bibinfo {author} {\bibfnamefont {J.~H.}\ \bibnamefont {Pixley}},
  \bibinfo {author} {\bibfnamefont {E.~J.}\ \bibnamefont {K{\"o}nig}},\ and\
  \bibinfo {author} {\bibfnamefont {R.-P.}\ \bibnamefont {Riwar}},\ }\bibfield
  {title} {\bibinfo {title} {Quasiperiodic circuit quantum electrodynamics},\
  }\href@noop {} {\bibfield  {journal} {\bibinfo  {journal} {arXiv preprint
  arXiv:2212.12382}\ } (\bibinfo {year} {2022})}\BibitemShut {NoStop}%
\bibitem [{\citenamefont {Herrig}\ \emph {et~al.}(2023)\citenamefont {Herrig},
  \citenamefont {Koliofoti}, \citenamefont {Pixley}, \citenamefont {König},\
  and\ \citenamefont {Riwar}}]{herrig2023emulating}%
  \BibitemOpen
  \bibfield  {author} {\bibinfo {author} {\bibfnamefont {T.}~\bibnamefont
  {Herrig}}, \bibinfo {author} {\bibfnamefont {C.}~\bibnamefont {Koliofoti}},
  \bibinfo {author} {\bibfnamefont {J.~H.}\ \bibnamefont {Pixley}}, \bibinfo
  {author} {\bibfnamefont {E.~J.}\ \bibnamefont {König}},\ and\ \bibinfo
  {author} {\bibfnamefont {R.-P.}\ \bibnamefont {Riwar}},\ }\href@noop {}
  {\bibinfo {title} {Emulating moir\'e materials with quasiperiodic circuit
  quantum electrodynamics}} (\bibinfo {year} {2023}),\ \Eprint
  {https://arxiv.org/abs/2310.15103} {arXiv:2310.15103 [cond-mat.mes-hall]}
  \BibitemShut {NoStop}%
\end{thebibliography}%

\end{document}